\documentclass[final,3p,12pt]{elsarticle}
\usepackage{graphicx}
\usepackage{amssymb}
\usepackage{amsmath}
\usepackage{relsize}
\biboptions{compress}
\journal{Annals of Physics}
\def\ket#1{|#1\rangle}
\def\bra#1{\langle#1|}
\def\scal#1#2{\langle#1|#2\rangle}
\def\matr#1#2#3{\langle#1|#2|#3\rangle}

\def\dirsum#1#2{\smallmatrix #2\\ \bigoplus \\ #1 \endsmallmatrix}
\def\dirprod#1#2{\smallmatrix #2\\ \bigotimes \\ #1 \endsmallmatrix}
\def\ave#1{\langle #1\rangle}

\def\jj1{j(j\!+\!1)}
\def\xx1#1#2{#1(#1\!#2\!1)}
\def\mpm1{m\!\pm\!1}
\def\={\!=\!}
\def\-{\!-\!}
\def\+{\!+\!}
\def\>{\!>\!}
\def\<{\!<\!}

\def\uvo#1{\lq\lq #1\rq\rq}
\def\uvo#1{\lq\lq #1\rq\rq}
\def\ESQPT{\textsc{esqpt}}
\def\ESQPTs{\textsc{esqpt}s}
\def\QPT{\textsc{qpt}}

\def\D{\textsc{d}}
\def\TC{\textsc{tc}}
\def\N{\textsc{n}}
\def\S{\textsc{s}}

\begin{document}

\begin{frontmatter}
\title{Quantum phases and entanglement properties \\ of an extended Dicke model}
\author[cuni]{Michal Kloc\corref{cor}}
\author[cuni]{Pavel Str{\'a}nsk{\'y}}
\author[cuni]{Pavel Cejnar}
\address[cuni]{Institute of Particle and Nuclear Physics, Faculty of Mathematics and Physics, Charles University, V~Hole{\v s}ovi{\v c}k{\'a}ch 2, Prague, 18000, Czech Republic}
\cortext[cor]{The corresponding author; email address: kloc@ipnp.troja.mff.cuni.cz}
\begin{abstract}
We study a simple model describing superradiance in a system of two-level atoms interacting with a single-mode bosonic field. The model permits a continuous crossover between integrable and partially chaotic regimes and shows a complex thermodynamic and quantum phase structure. Several types of excited-state quantum phase transitions separate quantum phases that are characterized by specific energy dependences of various observables and by different atom--field and atom--atom entanglement properties. We observe an approximate revival of some states from the weak atom--field coupling limit in the strong coupling regime. 
\end{abstract}
\end{frontmatter}

\section{Introduction}

Since its prediction in 1954 \cite{Dic54}, the effect of superradiance has attracted a lot of theoretical and experimental attention \cite{Gro82,Be96,Bra05}.
Its basic principle---the fact that a coherent interaction of an unexcited gas with the vacuum of a common field can create a spontaneous macroscopic excitation of both matter and field subsystems---appears in various incarnations in diverse branches of physics \cite{Bek98,Aue11}. 

The Dicke model \cite{Dic54,Jay63,Tav68} of superradiance resorts to a maximal simplification of the problem to capture the main features of the superradiant transition in the most transparent way. 
The model shows a thermal phase transition, analyzed and discussed in Refs.\,\cite{Wan73,Hep73a,Hep73b,Rza75}, as well as a zero-temperature (ground-state) Quantum Phase Transition (\QPT) \cite{Ema03a,Ema03b,Vid06}, which was addressed experimentally and realized with the aid of a superfluid gas in a cavity \cite{Dim07,Bau10,Bau11}.
Recent theoretical analyses showed that the model exhibits also a novel type of criticality---the one observed in the spectrum of excited states \cite{Per11a,Per11b,Bra13,Bas14}.
These so-called Excited-State Quantum Phase Transitions (\ESQPTs) affect both the density of quantum levels as a function of energy and their flow with varying control parameters, and are present in a wide variety of quantum models with low numbers of degrees of freedom \cite{Cej06,Cap08,Cej08,Str14,Str15,Str16}.

In this work, we analyze properties of a simple superradiance model interpolating between the familiar Dicke \cite{Dic54} and Tavis-Cummings \cite{Jay63,Tav68} Hamiltonians.
A smooth crossover between both limiting cases is achieved by a continuous variation of a model parameter, which allows one to observe the system's metamorphosis on the way from the fully integrable (hence at least partly understandable) to a partially chaotic (so entirely numerical) regime.
One of the aims of our work is to survey the phase-transitional properties of the extended model and to investigate the nature of its {\em quantum phases\,}---the domains of the excited spectrum in between the \ESQPT\ critical borderlines.
A usual approach to characterize different phases related to the ground state of a quantum system makes use of suitable \lq\lq order parameters\rq\rq, i.e., expectation values of some observables.
We show that an unmistakable characterization of phases involving excited states is not achieved through the expectation values alone but rather through their different smoothed energy dependences (trends).

The second part of our analysis is devoted to the {\em entanglement properties of excited states\/} across the whole spectrum and their potential links to the \ESQPTs\  and quantum phases of the model.
It is known that a continuous ground-state \QPT\ in many models (including the present one) is characterized by a singular growth of entanglement within the system, which can be seen as a quantum counterpart of the diverging correlation length in continuous thermal phase transitions \cite{Ost02,Vid04a,Vid04b,Lam04,Lam05,Bar06,Vid07,Bak12}.
A question therefore appears whether there exist any entanglement-related signatures of \ESQPTs.
The extended Dicke model is rather suitable for a case study of this type since it allows one to analyze at once various types of entanglement---that between the field and all atoms, and that between individual atoms.

The plan of the paper is as follows:
Basic quantum and classical features of the model are described in Sec.\,\ref{EDM}.
Thermal and quantum critical properties and a classification of thermodynamic and quantum phases are presented in Sec.\,\ref{PT}.
The atom--field and atom--atom entanglement properties are investigated in Sec.\,\ref{ENT}.
Conclusions come in Sec.\,\ref{SUM}.

\section{Extended Dicke model}
\label{EDM}
\subsection{Hamiltonian, eigensolutions, classical limit}
\label{HA}

Consider single-mode electromagnetic field with photon energy $\omega$ (polarization neglected) interacting with an ensemble of $N$ two-level atoms, all with the same level energies $\pm\omega_0/2$.
The size of the atomic ensemble is assumed to be much smaller than the wavelength of photons (cavity size) so that all atoms interact with the field with the same phase.
If we introduce an overall interaction strength $\lambda$ and an additional interaction parameter $\delta$ (whose role will be explained later), the Hamiltonian can be written as  
\begin{eqnarray}
H&=& 
\omega\,b^{\dag}b+\omega_0\sum_{k=1}^{N}\tfrac{1}{2}\sigma^k_z
+\frac{\lambda}{\sqrt{N}}\left[\left(b^{\dag}\+b\right)\sum_{k=1}^{N}\tfrac{1}{2}\left(\sigma^k_+\+\sigma^k_-\right)
-(1\-\delta)\sum_{k=1}^{N}\tfrac{1}{2}\left(b^{\dag}\sigma^k_+\+b\sigma^k_-\right)\right]
\nonumber\\
&=&
\underbrace{\omega\,b^{\dag}b+\omega_0J_z}_{H_{\rm free}}
+\frac{\lambda}{\sqrt{N}}
\underbrace{\left[b^{\dag}J_-+bJ_++\delta\ b^{\dag}J_++\delta\ bJ_-\right]}_{H_{\rm int}}
\,,
\label{H}
\end{eqnarray}
where operators $b^{\dag}$ and $b$ create and annihilate photons, while $\sigma_{\bullet}^k$ stands for the respective Pauli matrix with subscript $\{+,-,z\}$ or $\{x,y,z\}$ acting in the 2-state Hilbert space of the $k$th atom.
The part of $H$ denoted as $H_{\rm free}$ represents the free Hamiltonian of the field and atomic ensemble, while the part $H_{\rm int}$ constitutes the atom--field interaction with a conveniently scaled strength $\lambda/\sqrt{N}$.
Defining collective quasi-spin operators $J_{\bullet}=\sum_{k}\frac{1}{2}\sigma^k_{\bullet}$ for the whole atomic ensemble (which is possible due to the long wavelength assumption), we rewrite the whole Hamiltonian in the simplified form given in the second line of Eq.\,\eqref{H}.

The interaction part of the Hamiltonian \eqref{H} contains a parameter $\delta$.
For $\delta\=1$ we obtain the standard Dicke Hamiltonian \cite{Dic54}, in which
the interaction is written in the dipole approximation (the dipole operator for the $k$th atom is proportional to $\sigma^k_+\+\sigma^k_-\=2\sigma^k_x$ and the coupling strength in units of energy is given by $\lambda\=\omega_0dN^{1/2}/\epsilon_0\omega V^{1/2}$, with the electric dipole moment matrix element $d$, vacuum permitivity $\epsilon_0$ and cavity volume $V$).
This Hamiltonian is sometimes simplified by omitting the terms $b^{\dag}\sigma^k_+$ and $b\sigma^k_-$ that for very small $\lambda$ yield negligible contributions to the transition amplitudes \cite{Jay63,Tav68}.
The reduced model with $\delta\=0$, in case of $N\>1$ atoms called the Tavis-Cummings Hamiltonian \cite{Tav68}, conserves the sum of atomic and field excitation quanta and is integrable.
In this work, following Refs.\,\cite{Bra13,Bas14}, we analyze properties of an extended model across the transition between both the above limiting cases.
We assume that parameter $\delta$ in Eq.\,\eqref{H} varies smoothly within the interval $\delta\in[0,1]$, whose boundary values represent the Tavis-Cummings and Dicke Hamiltonians. 

The Hamiltonian \eqref{H} with any parameter setting conserves the squared quasi-spin $J^2\=J_x^2\+J_y^2\+J_z^2$ with eigenvalues $j(j\+1)$, where $j$ is integer for $N$ even or half-integer for $N$ odd \cite{Dic54}.
The full atomic Hilbert space ${\cal H}_{\rm A}$ is the span of all $2^N$ possible configurations of atoms, but due to the conservation of $J^2$ the dynamics can be investigated separately in any of the single-$j$ subspaces ${\cal H}_{\rm A}^{j,l}$ with dimension $2j\+1$.
The decomposition reads as follows \cite{Tav68}
\begin{equation}
{\cal H}_{\rm A}=\dirprod{k=1}{N}\underbrace{{\cal H}^k_{\rm A}}_{\mathbb{C}^2}
=\!\!\!\dirsum{j=0{\rm\,or\,}\frac{1}{2}}{\frac{N}{2}}
\!\!\left(\dirsum{l=1}{R_j}{\cal H}_{\rm A}^{j,l}\right)
\label{Hil}
\,,
\end{equation}
where $l$ enumerates replicas (their number is $R_j\=[N!(2j\+1)]/[(\frac{N}{2}\+j+1)!(\frac{N}{2}\-j)!]$) of the space with given $j$ differing by the exchange symmetry of atomic components.
In the following, we will investigate thermodynamic properties of the model in the full atomic space ${\cal H}_{\rm A}$, as well as quantum properties in a single-$j$ space ${\cal H}_{\rm A}^{j,l}$. 
The most natural choice in the latter case is the unique ($R_j$=1) subspace with maximal value $j\=N/2$, which is fully symmetric under the exchange of atoms and therefore emphasizes the collective character of the superradiance phenomenon.
A general-$j$ subspace has a mixed exchange symmetry such that only a number $N^*\=2j\leq N$ of atoms can be excited independently, while excitations of the remaining $N\-N^*$ atoms have to compensate each other (in Ref.\,\cite{Dic54} the quantum number $j$ is called a \uvo{cooperation number of the atomic gas}).
The reduced single-$j$ model has only two effective degrees of freedom $f$, one associated with the bosonic field, the other with the SU(2) algebra of collective quasi-spin operators, hence $f\=2$ \cite{Ema03a,Ema03b}.
In contrast, in the full (all-$j$) model the SU(2) algebra of Pauli matrices for each atom brings an independent degree of freedom, so the whole atom--field system has $f\=N\+1$.

The Tavis-Cummings Hamiltonian with $\delta\=0$ conserves the sum $M'=b^{\dag}b+J_z$ \cite{Tav68}.
For any fixed $j$, the conserved quantity can be written as
\begin{equation}
M=M'+j=\underbrace{b^{\dag}b}_{n}+\underbrace{J_z+j}_{n^*}
\label{M}
\,,
\end{equation}
where $n$ is the number of field bosons and $n^*$ the number of excited atoms (taking values $n^*\=m\+j\in[0,N^*]$, where $m$ is the $J_z$ quantum number).
The solutions of the $\delta\=0$ model are therefore restricted to any fixed-$M$ subspace ${\cal H}^{j,l}_{M}$ of the full Hilbert space 
\begin{equation}
{\cal H}={\cal H}_{\rm A}\otimes{\cal H}_{\rm F}\supset
{\cal H}^{j,l}_{\rm A}\otimes{\cal H}_{\rm F}=\dirsum{M=0}{\infty}{\cal H}^{j,l}_{M}
\label{HilM}
\,.
\end{equation}
Quantity \eqref{M} is not conserved in $\delta\neq 0$ cases, but the full Hamiltonian \eqref{H} always
conserves \uvo{parity}\ $\Pi\=(-)^M$ as the sum of atomic and field excitation quanta is varied only in pairs.

\begin{figure}[!t]
\includegraphics[width=\linewidth]{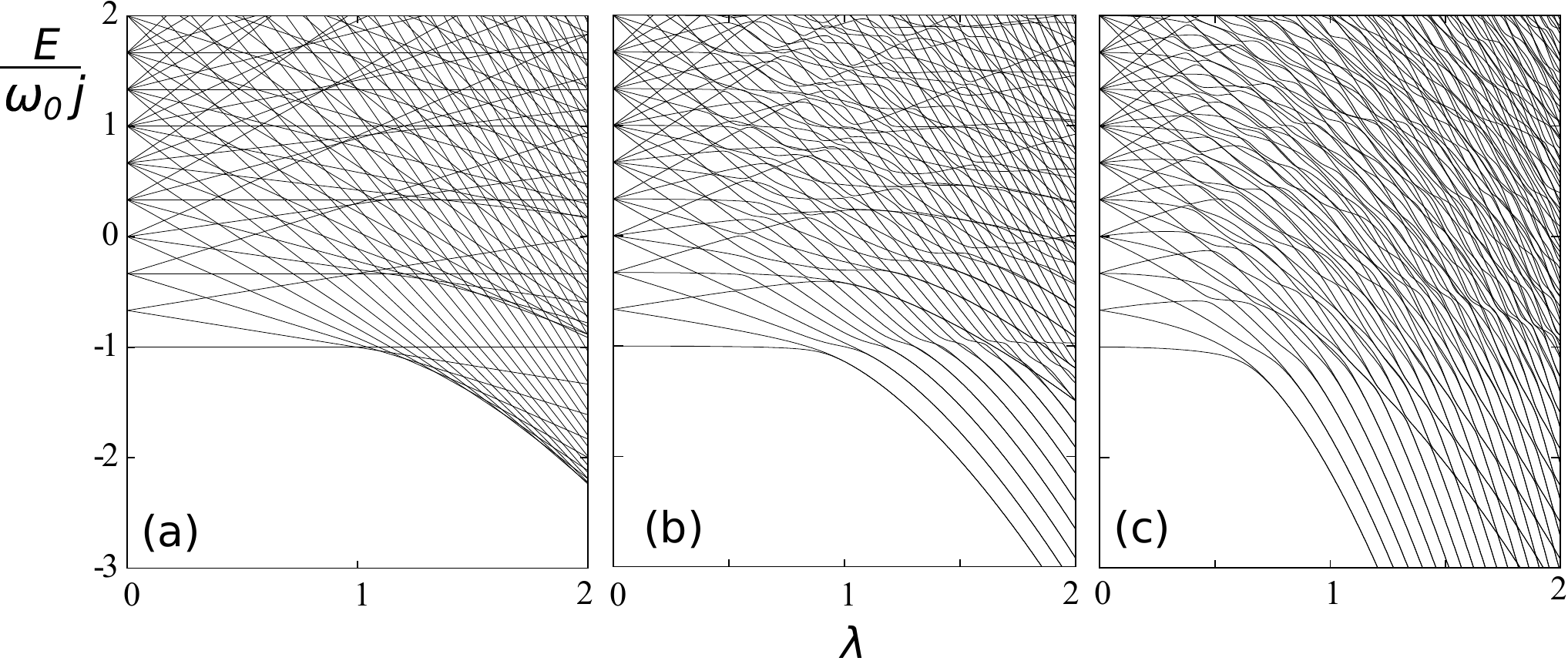}
\caption{Spectra of quantum energies of the $j\!=\!N/2$ extended Dicke model as a function of $\lambda$ for $\omega\!=\!\omega_0\!=\!1$ and $N\!=\!6$.
The panels correspond to (a) the Tavis-Cummings limit $\delta\!=\!0$, (b) intermediate regime $\delta\!=\!0.3$ and (c) the Dicke limit $\delta\!=\!1$.}
\label{f_spe}
\end{figure}

Figure~\ref{f_spe} shows quantum spectra of Hamiltonian \eqref{H} with $j\=N/2$ depending on the interaction parameter $\lambda$ for three different values of $\delta$.
The evolution of the ground state, namely its sudden drop to lower energies above a certain value of $\lambda$, indicates the superradiant transition at zero temperature.
We note that the eigensolutions of the model for a given $j$ are in general obtained by a numerical diagonalization of Hamiltonian \eqref{H} in the space ${\cal H}^{j,l}_{\rm A}\otimes{\cal H}_{\rm F}$ from Eq.\,\eqref{HilM}, using the basis $\ket{m}^{j,l}_{\rm A}\equiv\ket{m}_{\rm A}$ in ${\cal H}^{j,l}_{\rm A}$ (with $m\=-j,-j\+1,...,+j$) and $\ket{n}_{\rm F}$ in ${\cal H}_{\rm F}$ (with $n\=0,1,2,...$). 
Since the diagonalization procedure requires a truncation of ${\cal H}_{\rm F}$, the eigensolutions must be checked for convergence.

\begin{figure}[t!]
 \includegraphics[width=\linewidth]{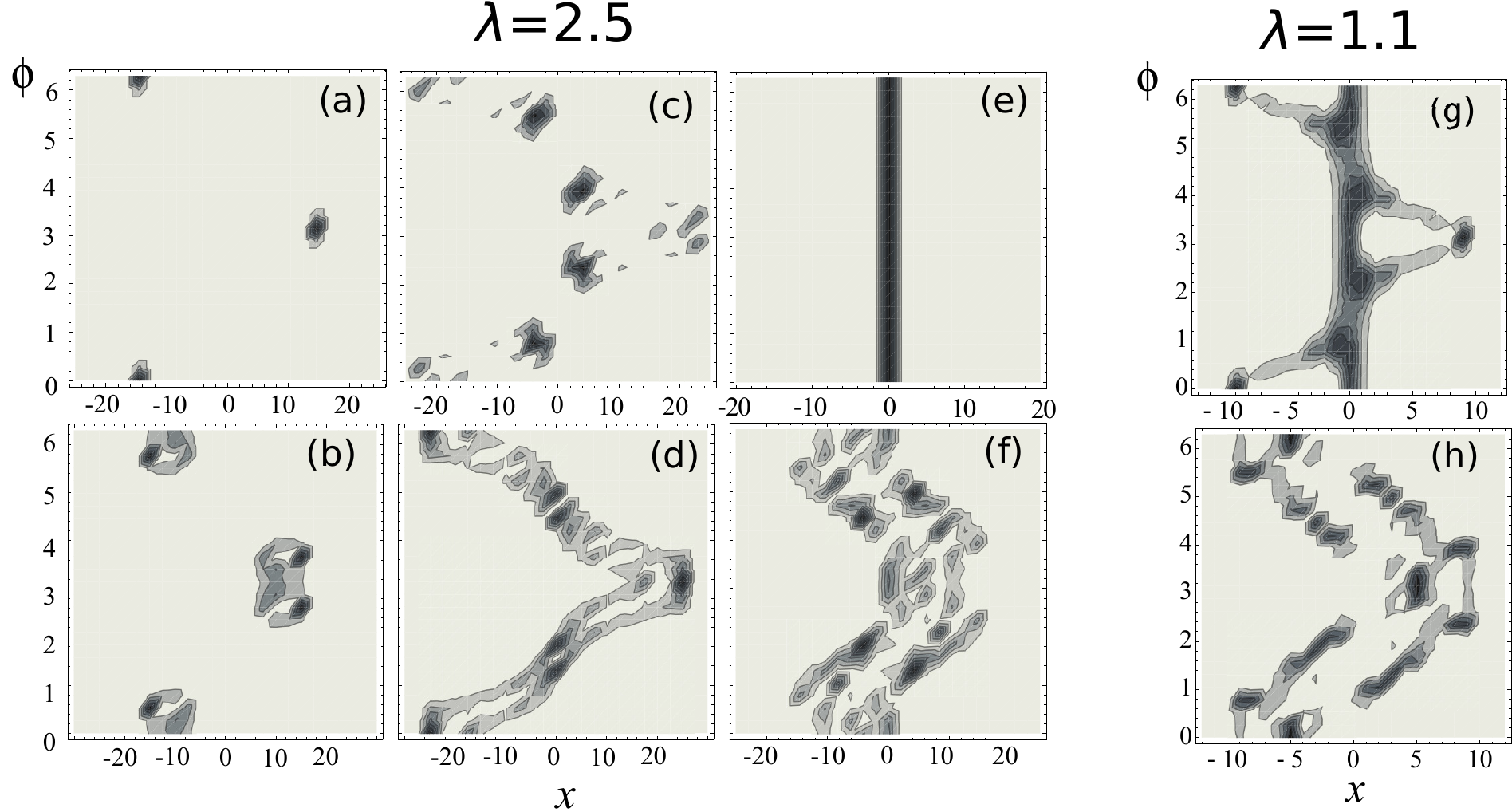}
\caption{ 
Squared wave functions $|\psi(\phi,x)|^2$ of selected eigenvectors of the Hamiltonian \eqref{H} with $N\!=\!40$, $\omega\!=\!\omega_0\!=\!1$ for $\delta\!=\!0.3$ taken at the $\lambda\times E$ values indicated by dots in Fig.\,\ref{f_qpd}d.
Panels (a)--(f), respectively, correspond to the vertical row of points at the right of Fig.\,\ref{f_qpd}d from bottom to top [panel (a) shows the ground state], panels (g),(h) to the left pair of points.
Note that variable $\phi$ is $2\pi$ periodic.
}
\label{f_wav}
\end{figure}

\begin{figure}[!t]
\includegraphics[width=\linewidth]{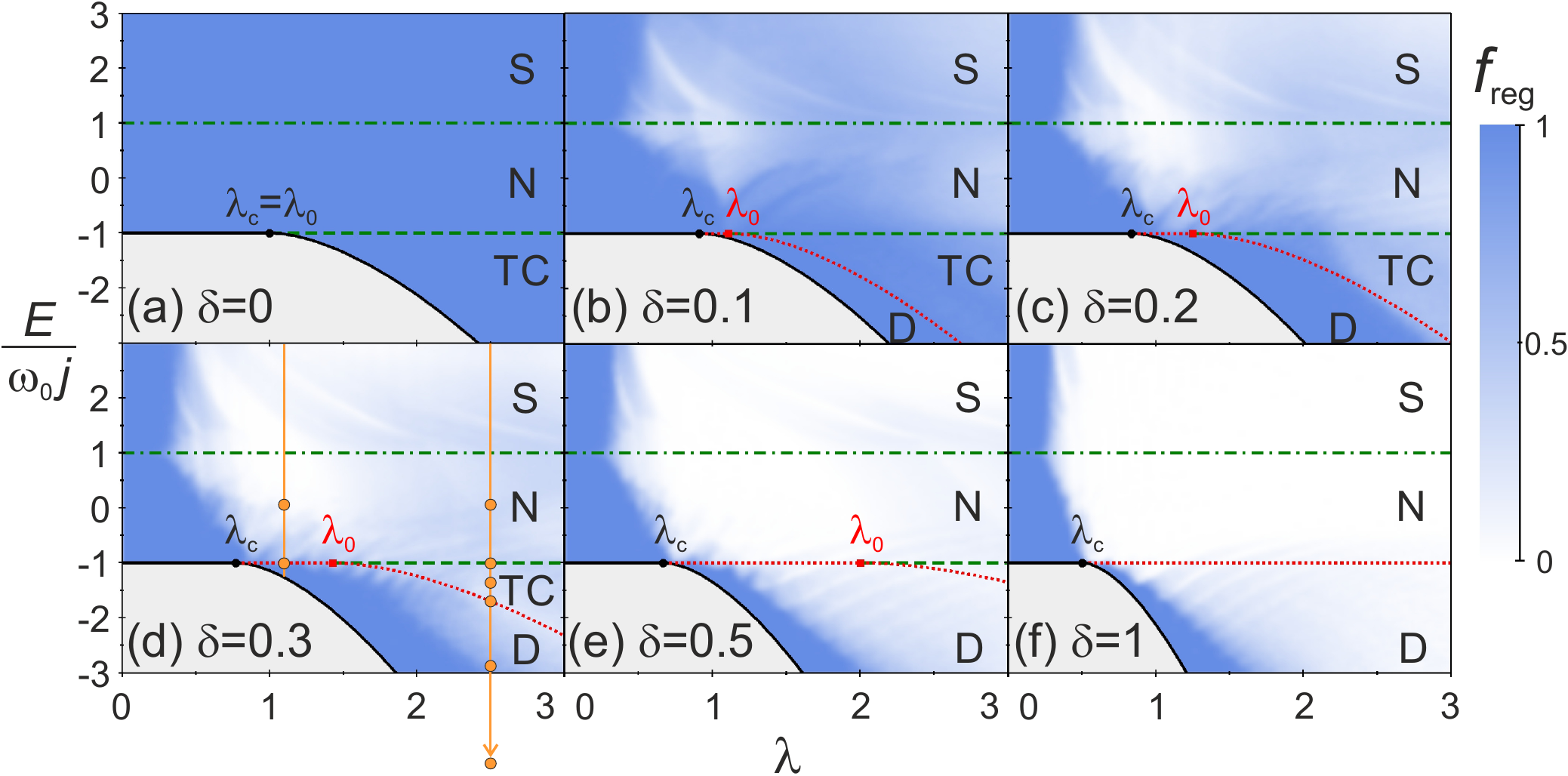}
\caption{(Color online)
The $\lambda\times E$ maps of the single-$j$ model with $\omega\!=\!\omega_0\!=\!1$ for six values of parameter~$\delta$.
Tones of blue indicate the classical regular fraction \eqref{freg}.
The curves correspond to the ground state energy $E_0$ (full line) and the critical \ESQPT\ energies $E_{c1}$ (dotted), $E_{c2}$ (dashed), and $E_{c3}$ (dash-dotted) from Eqs.\,(\ref{Ec1})--(\ref{Ec3}). 
Acronyms \D\ (Dicke), \TC\ (Tavis-Cummings), \N\ (normal) and \S\ (saturated) indicate different \lq\lq quantum phases\rq\rq\ of the model (Sec.\,\ref{QPT}).
Dots in panel (d) mark places where wave functions in Fig.\,\ref{f_wav} are taken. 
}
\label{f_qpd}
\end{figure}

A general state vector for any fixed $j$ is expressed as 
\begin{equation}
\ket{\psi}=\sum_{m=-j}^{+j}\sum_{n=0}^{\infty}\alpha_{mn}\ket{m}_{\rm A}\ket{n}_{\rm F}
\,,
\label{sta}
\end{equation}
where $\alpha_{mn}$ are expansion coefficients satisfying the normalization condition $\sum_{m,n}|\alpha_{mn}|^2\=1$.
Note that for even- and odd-parity states, respectively, the sums in Eq.\,\eqref{sta} go either over even or odd values of $M\=n\+m\+j$.
Any vector \eqref{sta} can be visualized as a wave function $\psi(\phi,x)$, where the angle $\phi\in[0,2\pi)$ comes from the quasi-spin representation of the atomic subsystem and variable $x\in(-\infty,+\infty)$ from the oscillator representation of the bosonic field.
For integer $j$ (even $N$), the wave function can be obtained by substitutions (up to normalization constants) $\ket{m}_{\rm A}\mapsto e^{im\phi}$ and $\ket{n}_{\rm F}\mapsto H_n(\omega^{1/2}x)e^{-\omega x^2/2}$, with $H_n$ standing for the Hermite polynomial.
For half-integer $j$ (odd $N$), however, the wave function $\psi(\phi,x)$ acquires a spinorial character.
A possible representation on the interval $\phi\in[0,2\pi)$ can be obtained by coupling an integer angular momentum $j'\=j\-1/2$ (the even core of $N\-1$ atoms) with $j''\=1/2$ (the odd atom) to the total $j$ as in the case of spinor spherical harmonics. 
This leads to a mapping of $\ket{m}_{\rm A}$ to a 2-valued function $\propto C_{\pm}e^{i(m\pm 1/2)\phi}$, which is $2\pi$-periodic, with $C_{\pm}$ denoting the corresponding Clebsch-Gordan coefficients.
Examples of squared wave functions of the $j\=N/2$ Hamiltonian eigenstates for an even $N$ in the medium-$\delta$ regime are depicted in Fig.\,\ref{f_wav}.
They are taken at the parameter and energy values indicated by dots in Fig.\,\ref{f_qpd}d ($\delta\=0.3$).
All eigenstates have a good parity, which means that their wave functions are symmetric or antisymmetric under the transformation $\phi\mapsto(\phi\-\pi)\,{\rm mod}\,2\pi$ and $x\mapsto -x$.

In the following, we will need the classical limit of the model.
It is constructed by a simple operator to c-number mappings according to (see, e.g., Ref.\,\cite{Bak13})
\begin{eqnarray}
(J_x,J_y,J_z)&\mapsto&j(\sin\theta\cos\phi,\sin\theta\sin\phi,-\cos\theta)
\label{map1},\\
(b,b^{\dag})&\mapsto&\tfrac{1}{\sqrt{2}}\bigl(x+ip,x-ip\bigr)
\label{map2}\,.
\end{eqnarray}
The spherical angles $\phi\!\in\![0,2\pi)$ and $\theta\!\in\![0,\pi]$ determine the orientation of the classical quasi-spin vector $\vec{j}\=(j_x,j_y,j_z)$.
Note that angle $\theta$ is measured here from the south to the north pole, thus a fully de-excited state of atoms ($j_z\=-j$) corresponds to $\theta\=0$ while a maximally excited state of atoms ($j_z\=+j$) is associated with $\theta\=\pi$.
It can be shown that the pair of variables $\phi$ and $j_z\=-j\cos\theta$, respectively, represents canonically conjugate coordinate and momentum associated with the collective states of the atomic subsystem.
Similarly, quantities $x\!\in\!(-\infty,+\infty)$ and $p\!\in\!(-\infty,+\infty)$ in Eq.\,(\ref{map2}) are suitable coordinate and momentum of the field subsystem.
Let us point out that while the atomic phase space for each $N$ is finite (it is a ball with radius $j$), the field phase space covers the whole plane, expressing the fact that the number of bosons is unlimited.
Note that an alternative route to the classical limit of the Dicke model was taken in Refs.\,\cite{Ema03a,Ema03b}.

The classical Hamiltonian resulting from application of Eqs.\,\eqref{map1} and \eqref{map2} in \eqref{H} reads as
\begin{equation}
H_\textnormal{cl}=\omega\,\frac{p^2\+x^2}{2}+\omega_0j_z+
\lambda\,\sqrt{\frac{2}{N}\left(j^2\-j_z^2\right)}
\biggl[(1\+\delta)x\cos{\phi}-(1\-\delta)p\sin{\phi}\biggr]
\label{Hcl}
\end{equation}
(see also Ref.\,\cite{Bas14}).
Since the classical limit coincides with $j,N\to\infty$, the above expression needs an appropriate scaling.
It is achieved by introducing size-independent quantities $H_{\rm cl}/2j$ and $j_z/2j$ together with $(x,p)/\sqrt{2j}$.
Their application in Eq.\,\eqref{Hcl} leads to an expression for scaled energy containing an effective coupling strength $\lambda_{\rm eff}\equiv\lambda\sqrt{2j/N}$, which increases with $j$ and reaches the bare $\lambda$ for the maximal $j\=N/2$.

Since the parameter $\delta$ drives the system away from the integrable Tavis-Cummings limit, it is relevant to measure the degree of chaos present at various stages of the transition to the Dicke limit. 
We note that chaos in the basic and extended Dicke model was studied also in Refs.\,\cite{Ema03a,Ema03b,Bas15,Bas16,Cha16}.
Classical chaos in a general Hamiltonian \eqref{Hcl} can be quantified by a so-called regular fraction $f_{\rm reg}$.
It is defined as the volume of the phase space domain filled with regular orbits relative to the total phase space volume available at given energy $E$:
\begin{equation}
f_{\rm reg}(E)=\frac{\int d\phi\,dj_z dx\,dp\ \delta(E\-H_\textnormal{cl})\ \chi_{\rm reg}
}{\int d\phi\,dj_z dx\,dp\ \delta(E\-H_\textnormal{cl})}
\label{freg}
\,.
\end{equation}
Here, $\chi_{\rm reg}(\phi,j_z,x,p)$ stands for a characteristic function defining the regular part of the phase space, i.e., the domain where classical dynamics is regular in the sense of vanishing Lyapunov exponents ($\chi_{\rm reg}\=1$ in the regular part and $\chi_{\rm reg}\=0$ in the chaotic part).
The way how the regular fraction is obtained from numerical simulation of classical dynamics is described in Ref.\,\cite{Cej14}.

The evolution of the classical regular fraction for Hamiltonian \eqref{Hcl} with $\delta$ increasing from 0 to 1 is shown in Fig.\,\ref{f_qpd}.
Each panel displays a map of $f_{\rm reg}$ for a given $\delta$ in the plane $\lambda\times E$.
The value of $f_{\rm reg}$ is encoded into the tones of blue: the full color indicates perfectly regular areas ($f_{\rm reg}\=1$) and white completely chaotic ones ($f_{\rm reg}\=0$).
Not surprisingly, we observe that the degree of chaos exhibits an overall increase with $\delta$.
However, even in the Dicke limit $\delta\=1$ the model is not entirely chaotic.
The most chaotic domain in all $\delta\>0$ panels is that with parameter $\lambda$ above the superradiant transition and energy $E$ exceeding a certain value above the ground-state $E_0$.
Besides the main trends in the dependence of $f_{\rm reg}$ we also observe some surprising fine structures---for instance the \lq\lq ribs\rq\rq\ in panels (d)--(f).

\subsection{Classical and quantum solutions with fixed $M$ for $\delta=0$}
\label{INT}

\begin{figure}[t!]
\includegraphics[width=0.85\linewidth]{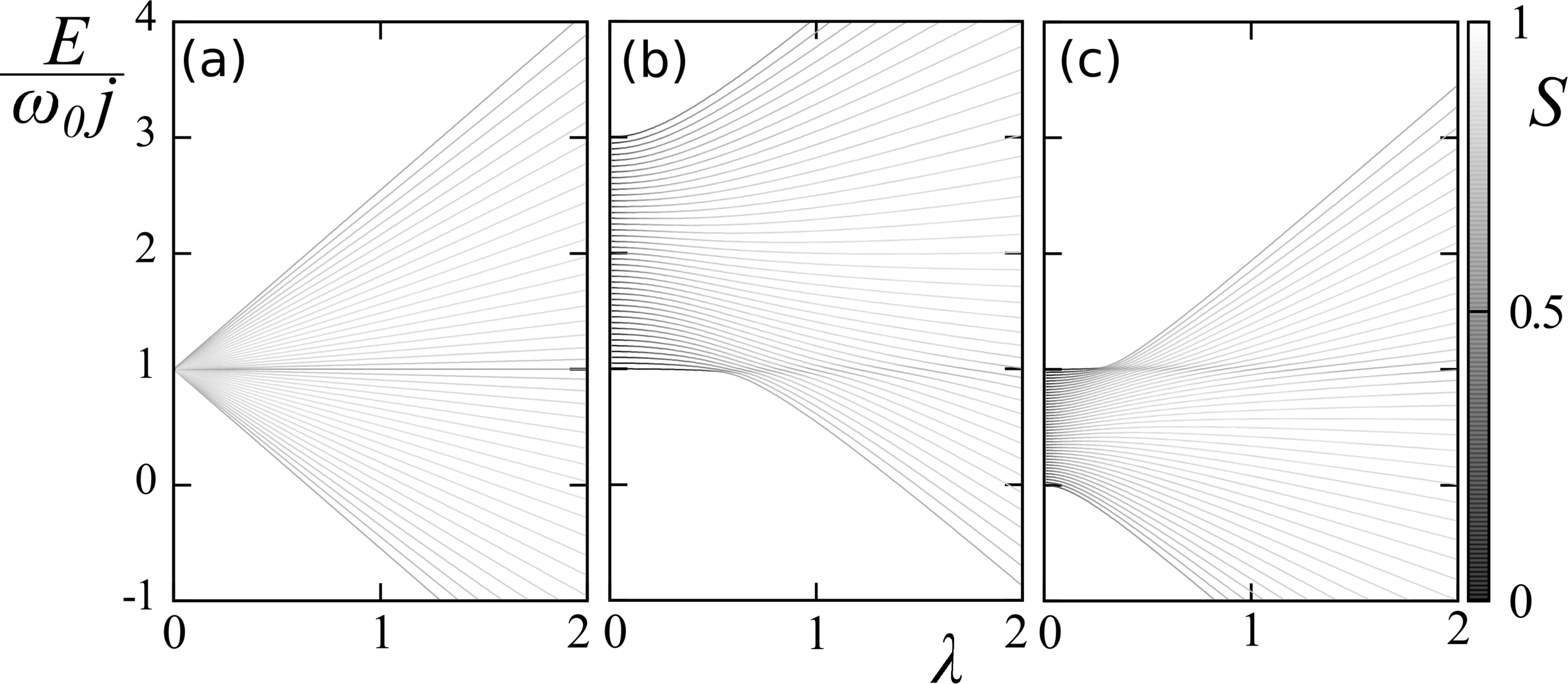}
\caption{
The $j\!=\!N/2$ spectrum (for $N\!=\!40$) in the $M\!=\!2j$ subspace for the $\delta\!=\!0$ model with $\omega\!=\!\omega_0\!=\!1$ (panel a), $\omega\!=\!2,\omega_0\!=\!1$ (panel b) and $\omega\!=\!0.5,\omega_0\!=\!1$ (panel c). The wave-function entropies ${\cal S}$ of individual eigenstates in the unperturbed basis $\ket{m}_{\rm A}\ket{n}_{\rm F}$ are encoded in the varying shade of the corresponding lines.}
\label{f_cary}
\end{figure}

Finding eigensolutions of Hamiltonian \eqref{H} is much simpler in the $\delta\=0$ limit \cite{Tav68}. 
In that case, the additional conserved quantity \eqref{M} splits the Hilbert space ${\cal H}^{j,l}_{\rm A}\otimes{\cal H}_{\rm F}$ from Eq.\,\eqref{HilM} into a sum of dynamically invariant subspaces ${\cal H}^{j,l}_M$.
These are spanned by states $\ket{m}_{\rm A}\ket{n}_{\rm F}$ satisfying constraint $n+m=M-j$ and therefore have dimension $d(j,M)\={\rm min}\{2j,M\}+1$.
As seen in Fig.\,\ref{f_spe}a, the full $\delta\=0$ spectrum is comprised of a number of mutually non-interacting (crossing each other) spectra with different values of $M$. 
Each of these fixed-$M$ spectra is obtained by truncation-free diagonalization in a finite dimension.
Using the unperturbed basis $\ket{i}\equiv\ket{m\=-j\+i}_{\rm A}\ket{n\=M\-i}_{\rm F}$ enumerated by $i\=0,\!...,{\rm min}\{2j,M\}$, we express the $\delta\=0$ Hamiltonian in a single $M$-subspace in the following tridiagonal form:
\begin{equation}
\matr{i}{H}{i'}=\underbrace{\left[\omega(M\-i)+\omega_0(i\-j)\right]}_{E_i(0)}\delta_{ii'}
+\lambda
\underbrace{\sqrt{\frac{(i\+1)(2j\-i)(M\-i)}{N}}\left[\delta_{i(i'-1)}+\delta_{(i-1)i'}\right]}_{H'_{ii'}}
\,.
\label{Hamoun}
\end{equation}
Examples of single-$M$ spectra obtained from this Hamiltonian are given in Fig.\,\ref{f_cary}.

The simplest spectrum shown in Fig.\,\ref{f_cary}a corresponds to the {\em tuned\/} case with $\omega\=\omega_0$.
The eigenvalues of the tuned Hamiltonian \eqref{Hamoun} read as $E_i\=E(0)\+\lambda E'_i$, where $E(0)\=\omega(M\-j)$ and $E'_i$ are eigenvalues of $H'_{ii'}$.
Since the latter come in pairs with opposite signs (for odd dimensions there is an additional unpaired eigenvalue $E'_i\=0$), the spectrum linearly expands with increasing $\lambda$ around $E(0)$.
The full spectrum in Fig.\,\ref{f_spe}a results just from a pile up of spectra similar to that from Fig.\,\ref{f_cary}a with all values of $M$.
Moreover, as the eigenvectors of matrix \eqref{Hamoun} with $\omega\=\omega_0$ coincide with those of $H'_{ii'}$, they do not depend on $\lambda$. 
This is also verified in Fig.\,\ref{f_cary}a, where the wave-function entropy ${\cal S}(\lambda)$ of individual eigenstates in the unperturbed basis is encoded in the shade of the respective line.
The constancy of ${\cal S}(\lambda)$ for each state confirms the absence of its structural evolution.
The wave-function entropy for a general state \eqref{sta} with respect to the basis $\ket{m}_{\rm A}\ket{n}_{\rm F}$ is defined by the expression
\begin{equation}
{\cal S}(\psi)=-\tfrac{1}{\ln(2j+1)}\sum_{m,n}|\alpha_{mn}|^2\ln|\alpha_{mn}|^2
\label{wentro}
\end{equation}
(cf.\,Ref.\,\cite{Cej98}), so it measures the degree of delocalization of the given state in the fixed basis.
The minimal value ${\cal S}\=0$ is assigned to any of the basis states (no delocalization), while the maximum ${\cal S}\=\ln d(j,M)/\ln(2j\+1)$ is taken by a uniformly distributed superposition of all basis states (maximal delocalization).
In Sec.\,\ref{ENT} we will show that Eq.\,\eqref{wentro} quantifies the atom--field entanglement contained in the given state for $\delta\=0$.

The spectrum of a {\em detuned\/} Hamiltonian with $\omega\!\neq\!\omega_0$ is less trivial.
Examples of level dynamics in a single-$M$ subspace and the corresponding wave-function entropies for $\omega\>\omega_0$ and $\omega\<\omega_0$ are shown in panels (b) and (c) of Fig.\,\ref{f_cary}. 
We see that the eigenvalues are no more straight lines and that the eigenvectors change with $\lambda$.
The peculiar shapes of these spectra will be explained in Sec.\,\ref{QPT}, here we only point out that the tuned situation is restored in an approximate sense for very large $\lambda$, when the last term in the detuned Hamiltonian $H=\omega(n\+J_z)\-\lambda H_{\rm int}/\sqrt{N}\-(\omega\-\omega_0)J_z$ becomes negligible.
Therefore, investigating properties of the model for large coupling strengths, one can make an assumption that $\omega\!\approx\!\omega_0$.

Let us finally return to the classical analysis, now focused specifically on the $\delta\=0$ system.
The conservation of quantity $M$ from Eq.\,\eqref{M} introduces particular correlations between the degrees of freedom associated with atomic and field subsystems.
As a consequence, the system with $f\=2$ degrees of freedom is for any fixed value of $M$ reduced to a system with just a single effective degree of freedom.
To make this explicit, we employ a transformation
\begin{equation}
\left(\begin{array}{c}
x\\p\\\phi\\j_z
\end{array}\right)
\mapsto
\left(\begin{array}{l}
x'=x\cos\phi-p\sin\phi\\
p'=p\cos\phi+x\sin\phi\\
\phi'=\phi+j_z+(p^2\+x^2)/2\\
M'=j_z+(p^2\+x^2)/2
\end{array}\right)
\label{canon}
\,,
\end{equation}
where $M'\=M\-j$ is conserved.
It can be checked that this transformation is canonical, so $(x',p')$ and $(\phi',M')$ are new pairs of conjugate coordinates and momenta.
The $(x',p')$ variables for each $M'$ satisfy a constraint $(x'^2\+p'^2)/2\in[{\rm max}\{0,M'\-j\},M'\+j]$, so they form a disc (for $M'\leq j$) or an annulus (for $M'\>j$).
The classical Tavis-Cummings ($\delta\=0$) Hamiltonian in terms of the new variables becomes
\begin{equation}
H^\textnormal{TC}_\textnormal{cl}=
(\omega\-\omega_0)\,\frac{p'^2\+x'^2}{2}+\omega_0 M'
+\lambda\,x'\sqrt{\frac{2}{N}\left[j^2-\left(M'\-\frac{p'^2\+x'^2}{2}\right)^2\right]}
\label{Hcl1D}
\,,
\end{equation}
which for a constant $M'$ depends only on $(x',p')$, so has effectively $f\=1$ degree of freedom.
This not only guarantees the full integrability of the $\delta\=0$ model, but also results in some emergent critical phenomena which will be discussed below.

\section{Phases and phase transitions}
\label{PT}
\subsection{Thermal phase transition}
\label{TPT}

The extended Dicke Hamiltonian (\ref{H}) has a complex phase structure \cite{Bra13,Bas14,Bas16b}.
In this section, we analyze thermal phase transition of the full model with atomic Hilbert space (\ref{Hil}) and the number of degrees of freedom $f\=N\+1$.
Following the standard approach described in Ref.\,\cite{Wan73}, we use the Glauber coherent states $\ket{\alpha}\propto e^{\alpha b^{\dag}}\ket{0}$, where $\alpha\equiv(\alpha'\+i\alpha'')\in{\mathbb C}$ is an eigenvalue of the $b$ operator and $\ket{0}$ denotes the field vacuum.
An average number of bosons $\ave{n}_{\alpha}\=\matr{\alpha}{b^{\dag}b}{\alpha}\=|\alpha|^2$ in the state $\ket{\alpha}$ represents an indicator of the transition from normal to superradiant phase: $\ave{n}_{\alpha}\=0$
in the normal phase and $\ave{n}_{\alpha}\!\sim\!{\cal O}(N)$ in the superradiant phase. 
On the way to the thermodynamic limit $N\!\to\!\infty$, it is convenient to perform the scaling transformation $\alpha\mapsto{\bar\alpha}\=\alpha/\sqrt{N}$, so that the scaled coherent-state variable ${\bar\alpha}\equiv{\bar\alpha'}\+i{\bar\alpha''}$ becomes a suitable complex order parameter of the superradiant phase transition.

We start by introducing two special values of the atom--field coupling strength:
\begin{equation}
\lambda_c=\frac{\sqrt{\omega\omega_0}}{1+\delta}
\,,\quad
\lambda_0=\frac{\sqrt{\omega\omega_0}}{1-\delta}
\label{lc0}
\,.
\end{equation}
While the first one will be shown to represent a critical strength for the occurrence of the superradiant phase, the second value will turn out to be a strength at which a specific form of the superradiant phase occurs for $\delta\in(0,1)$.
To get an equilibrium field configuration at a given temperature $T$ (measured in energy units), we minimize in variable ${\bar\alpha}$ a scaled free energy $F_{{\bar\alpha}}(T)/N=-T\ln Z_{{\bar\alpha}}(T)/N$, where $Z_{{\bar\alpha}}(T)=\matr{{\bar\alpha}}{{\rm Tr}_{\rm A} e^{-H/T}}{{\bar\alpha}}$ is the partition function, calculated for a given ${\bar\alpha}$ from a partial trace of the Hamiltonian exponential over the full atomic Hilbert space (\ref{Hil}).
For Hamiltonian (\ref{H}) this procedure yields
\begin{equation}
Z_{{\bar\alpha}}(T)=e^{-\frac{\omega|{\bar\alpha}|^2}{T}}
{\rm Tr}^N{\rm exp}
\left\{
-\frac{1}{T}
\left(\begin{array}{cc}
\frac{1}{2}\omega_0 & \lambda\left[(1\+\delta){\bar\alpha'}\!\+i(1\-\delta){\bar\alpha''}\right] \\
\lambda\left[(1\+\delta){\bar\alpha'}\!\-i(1\-\delta){\bar\alpha''}\right] & \frac{1}{2}\omega_0
\end{array}\right)
\right\}
\label{Z}
\,.
\end{equation}
The two eigenvalues of the matrix in the exponential are
\begin{equation}
\pm\sqrt{\left(\tfrac{\omega_0}{2}\right)^2\!\+\lambda^2\bigl[ (1\+\delta)^2 {\bar\alpha}^{\prime 2}\+ (1\-\delta)^2 {\bar\alpha}^{\prime\prime 2}\bigr]}\equiv\pm e_{{\bar\alpha}}
\label{e}
\,,
\end{equation}
which leads to a simple expression of the scaled free energy:
\begin{equation}
\tfrac{1}{N}F_{{\bar\alpha}}(T)=\omega|{\bar\alpha}|^2-T\ln\left(2\cosh\tfrac{e_{{\bar\alpha}}}{T}\right)
\label{F}
\,.
\end{equation}

Stationary points of $F_{{\bar\alpha}}(T)/N$ can be obtained from a simple analysis.
The first derivative in variable ${\bar\alpha}'$ or ${\bar\alpha}''$, respectively, vanishes at the points satisfying the first or second line of the following array:
\begin{eqnarray}
{\bar\alpha}'=0
&{\rm or}&
2\omega\,e_{{\bar\alpha}}=\lambda^2(1\-\delta)^2\tanh\tfrac{e_{{\bar\alpha}}}{T}
\label{con1}
\,,
\\
{\bar\alpha}''=0
&{\rm or}&
2\omega\,e_{{\bar\alpha}}=\lambda^2(1\+\delta)^2\tanh\tfrac{e_{{\bar\alpha}}}{T}
\label{con2}
\,.
\end{eqnarray}
The point ${\bar\alpha}\=0$ is always a trivial solution of both lines, but additional solutions appear for $\lambda\>\lambda_c$.
This coupling strength represents a critical point where the normal phase ${\bar\alpha}\=0$ becomes unstable at the lowest temperatures and a new, superradiant equilibrium is created at some non-zero values of ${\bar\alpha}$.
The critical temperature for the superradiant transition, i.e., the upper temperature limit for the existence of the ${\bar\alpha}\!\neq\!0$ solution in the region $\lambda\>\lambda_c$, is
\begin{equation}
T_c=\frac{\omega_0}{2}\,{\rm artanh}^{-1}\frac{\lambda_c^2}{\lambda^2}
\label{Tc}
\,.
\end{equation}
For $T\>T_c$, the stable equilibrium of the system is again only the normal solution ${\bar\alpha}\=0$.

\begin{figure}[!t]
\includegraphics[width=\linewidth]{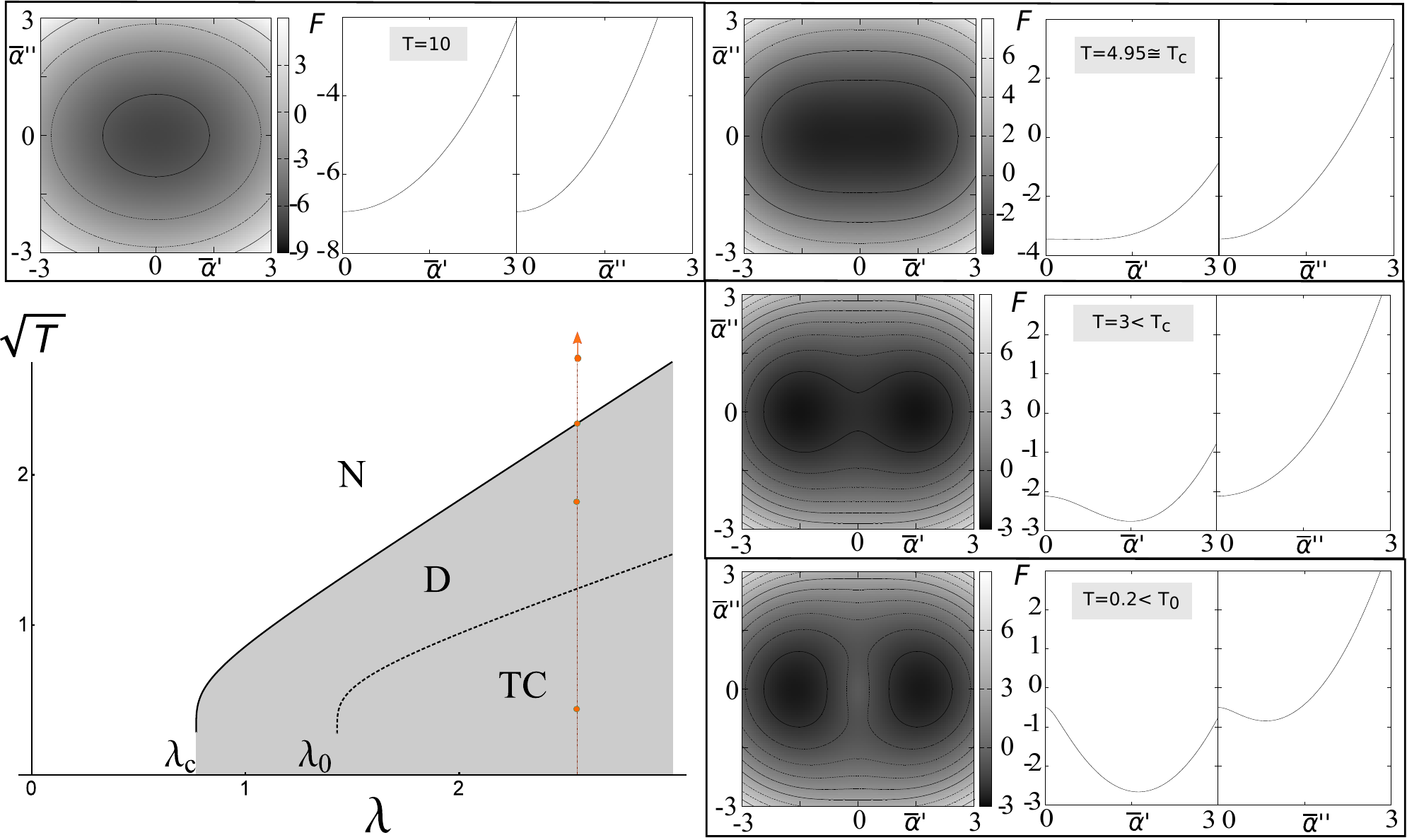}
\caption{
Thermal phase diagram of the all-$j$ model ($\omega\!=\!\omega_0\!=\!1$, $\delta\!=\!0.3$) and sample landscapes of the scaled free energy at the $\lambda\!=\!2.5$ cut of the phase diagram (vertical line). The full curve in the phase diagram marks the critical temperature $T_c$ for the transition to the normal (\N) phase, while the dotted curve indicates the temperature $T_0$ separating the \TC\ and \D\ superradiant phases (with or without saddles of free energy). The free energy landscapes in the complex-$\alpha$ plane are, for selected temperatures, visualized by contour-shade plots (darker areas indicate lower values) and by the cuts along real and imaginary axes.
}
\label{f_tpd}
\end{figure}

For $\delta\=0$ (the Tavis-Cummings limit), both right-side conditions in Eqs.\,(\ref{con1}) and (\ref{con2}) become identical and yield a solution $|{\bar\alpha}|\={\rm const}$ that grows from zero with increasing difference $\lambda\-\lambda_c$.
Therefore, the superradiant minimum of the free energy forms a circle around ${\bar\alpha}\=0$.
For $\delta\>0$, however, a simultaneous solution of both right-side conditions in (\ref{con1}) and (\ref{con2}) is no more possible.
The superradiant equilibrium is then represented by a pair of points on the real axis, $({\bar\alpha}',{\bar\alpha}'')\=(\pm{\bar\alpha}'_0,0)$, where ${\bar\alpha}'_0\>0$ solves the right-side equation in (\ref{con1}).
If $\delta\<1$ and $\lambda\>\lambda_0$, a pair of saddle points appears for low temperatures on the imaginary axis, $({\bar\alpha}',{\bar\alpha}'')\=(0,\pm{\bar\alpha}''_0)$, where ${\bar\alpha}''_0\geq {\bar\alpha}'_0$ solves the right-side equation in (\ref{con2}).
These unstable solutions exist for temperatures below
\begin{equation}
T_0=\frac{\omega_0}{2}\,{\rm artanh}^{-1}\frac{\lambda_0^2}{\lambda^2}
\label{T0}
\,.
\end{equation}
A possibility of thermodynamic quasi-equilibrium states associated with the saddle points of free energy \eqref{F} was recently discussed in Ref.\,\cite{Bas16b}.

A thermal phase diagram in the $\lambda\times T$ plane for $\delta\=0.3$ is shown in Fig.\,\ref{f_tpd} along with samples of the free energy landscapes in various phases and at $T\=T_c$.
The critical temperature $T_c$ from Eq.\,(\ref{Tc}) determines the phase transition between the superradiant and normal (acronym \N) phases of the model.
The superradiant phase exists in two forms: the Tavis-Cummings phase (acronym \TC) with the saddles in the free energy landscape and the Dicke phase (acronym \D) without the saddles.
While the \TC\ phase is the only type of superradiant phase in the Tavis-Cummings limit (where $\lambda_0\!\to\!\lambda_c$ and $T_0\!\to\!T_c$), the \D\ phase is exclusive in the Dicke limit (in which $\lambda_0\!\to\!\infty$).
For intermediate $\delta$ both phases coexist, being separated by temperature $T_0$ from Eq.\,(\ref{T0}).
Let us stress that the \TC$\,\to\,$\D\ transition (in contrast to \D$\,\to\,$\N) is not a phase transition in the standard sense since it does not affect the global minimum of the free energy.

\subsection{Ground-state and excited-state quantum phase transitions}
\label{QPT}

We return now to the analysis of the extended Dicke model with the restriction to a single-$j$ collective subspace of atomic states. 
The number of relevant degrees of freedom of the single-$j$ model is $f\=2$, independently of the size parameter $N$, which implies that the infinite-size limit, $N\to\infty$, coincides with the classical limit \cite{Per11a,Str14}.
Both ground-state and excited-state quantum phase transitions can be predicted from the classical version of the model, namely from the behavior of stationary points of the classical Hamiltonian \eqref{Hcl}.
In particular, the \ESQPTs\  result from singularities (non-analyticities) in the semiclassical density of states
\begin{equation}
\varrho_\textnormal{cl}(E)=
\frac{\partial}{\partial E}\frac{1}{(2\pi)^2}\int d\phi\,dj_z dx\,dp\ \Theta(E-H_\textnormal{cl})
\label{rhocl}
\,,
\end{equation}
where $\Theta(x)$ is the step function ($\Theta\=0$ for $x\<0$ and $\Theta\=1$ for $x\!\geq\! 0$), that appear at the points satisfying $\nabla H_\textnormal{cl}\=0$ (with $\nabla$ standing for the gradient in the phase space).

It has been shown \cite{Str16} that the \ESQPTs\  caused by non-degenerate stationary points---those which are locally quadratic as their Hessian matrix of second derivatives has only non-zero eigenvalues---can be classified by a pair of numbers $(f,r)$, where a so-called index of the stationary point $r$ is a number of negative eigenvalues of the Hessian matrix. 
In particular, for $f=2$, the first derivative of the level density in a vicinity of a stationary-point energy $E_{c}$ behaves as  
\begin{equation}
\frac{\partial\varrho_\textnormal{cl}}{\partial E}\propto
\left\{\begin{array}{ll}
(-)^{r/2}\Theta(E-E_{c})  & {\rm for\ } r=0,2,4 \\
(-)^{(r+1)/2}\ln|E-E_{c}| & {\rm for\ } r=1,3
\end{array}\right.
\label{2Dclas}
\end{equation}
hence exhibits either a jump (for $r$ even) or a logarithmic divergence (for $r$ odd) at the critical energy $E_{c}$.

\begin{figure}[!t]
\includegraphics[width=\linewidth]{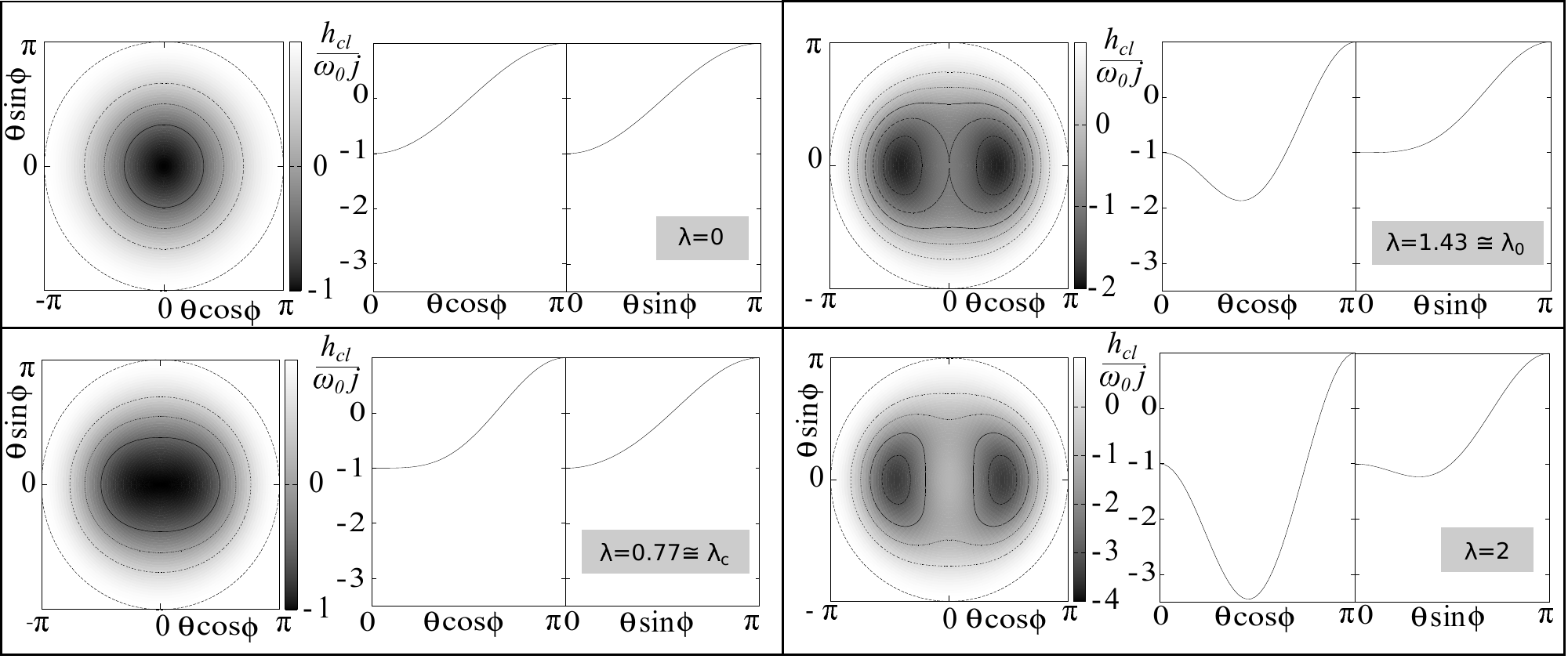}
\caption{The function $h_\textnormal{cl}$ from Eq.\,(\ref{trans}) in the phase space of the atomic subsystem defined by spherical angles $\theta$ and $\phi$ for $\omega\!=\!\omega_0\!=\!1$ and $\delta\!=\!0.3$.
Various panels correspond to the indicated values of the coupling strength $\lambda$.
Each contour-shade polar plot (with $\phi\equiv$\,angle and $\theta\equiv$\,radius) is accompanied by the corresponding horizontal and vertical cuts.
}
\label{f_impo}
\end{figure}

In the present case, the evaluation of the level density according to Eq.\,\eqref{rhocl} can be simplified with the aid of the volume-preserving substitution
\begin{equation}
\left(\begin{array}{c}x\\p\end{array}\right)
\mapsto 
\left(\begin{array}{c}
\xi=x+\frac{(1+\delta)\lambda}{\omega}\sqrt{\frac{2}{N}\left(j^2-j_z^2\right)}\cos\phi\\
\eta=p-\frac{(1-\delta)\lambda}{\omega}\sqrt{\frac{2}{N}\left(j^2-j_z^2\right)}\sin\phi
\end{array}\right)
\label{subs}
\,,
\end{equation}
which transforms the classical Hamiltonian \eqref{Hcl} to the form  
\begin{equation}
\ H_{\rm cl}'=\omega\,\frac{\eta^2+\xi^2}{2}+
\underbrace{
\omega_0j_z-\frac{\lambda^2}{\omega}\,\frac{j^2-j_z^2}{N}
\left(1+2\delta\cos 2\phi+\delta^2\right)
}_{h_{\rm cl}(\phi,j_z)}
\label{trans}
\end{equation}
with fully separated variables of the field and atomic subsystems. 
Although the mapping \eqref{subs} does {\em not\/} represent a canonical transformation (so $\xi,\eta$ is not a new coordinate-momentum pair---otherwise the model would be separable and thus fully integrable for any $\delta$), it simplifies the analysis of the level density.
In particular, the integration in Eq.\,\eqref{rhocl} over variables $\xi,\eta$, on which the transformed Hamiltonian depends quadratically, can be performed explicitly (cf. Ref.\,\cite{Str14}, where an analogous calculation is performed for a Hamiltonian with a quadratic kinetic term).
The result is a simplified expression
\begin{equation}
\varrho_\textnormal{cl}(E)=\frac{1}{2\pi\omega}\int d\phi\,dj_z\ \Theta(E-h_\textnormal{cl})
\label{rhocl2}
\,,
\end{equation}
in which the integration, involving function $h_{\rm cl}(\phi,j_z)$ defined in Eq.\,\eqref{trans}, goes only over the 2-dimensional collective phase space of the atomic subsystem---a ball with radius $j$.
Eq.\,\eqref{rhocl2} is proportional to an area of the ball region where $h_{\rm cl}$ takes values less than (or equal to) the chosen energy $E$, hence it can be visualized as flooding of a landscape with profile $h_{\rm cl}$ on a globe.
The $h_{\rm cl}$ function for selected values of $\lambda$ and $\delta$ is depicted in Fig.\,\ref{f_impo}.

\begin{figure}[!t]
\includegraphics[width=0.75\linewidth]{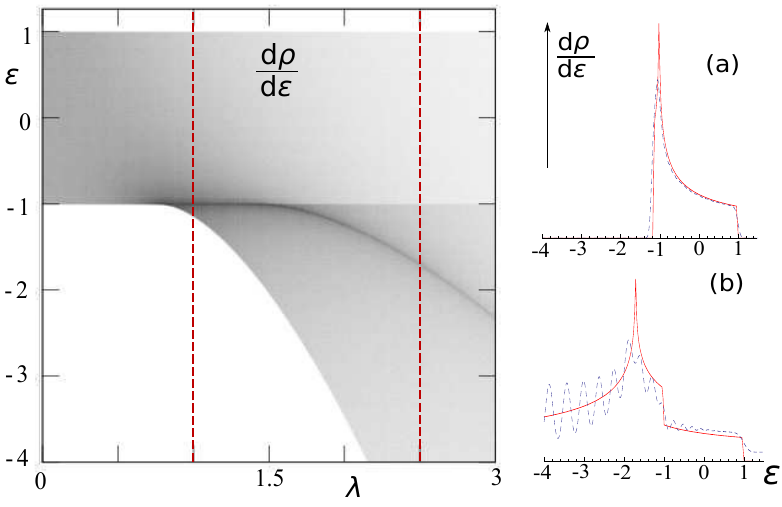}
\caption{Energy derivative of the semiclassical level density, $\partial\varrho_\textnormal{cl}/\partial E$, for the single-$j$ model with $\omega\!=\!\omega_0\!=\!1$ and $\delta\!=\!0.3$ as a function of $\lambda$ and $\varepsilon\!\equiv\!E/(\omega_0 j)$. 
The shade diagram in the left panel was obtained from Eq.\,\eqref{rhocl2}; darker areas represent larger values and vice versa.
Panels (a) and (b) show cuts (full curves) at (a) $\lambda\!=\!1$ and (b) $\lambda\!=\!2.5$ in comparison with finite-size results (dashed curves) based on numerical diagonalization of the Hamiltonian for $N\!=\!40$ and Gaussian smoothening of the spectrum ($\sigma\!=\!0.04$ and $0.07$ for $\lambda\!=\!1$ and $2.5$, respectively).
Scales in panels (a),(b) are arbitrary and not the same.}
\label{f_dender}
\end{figure}

It is clear from the expression \eqref{trans} that the stationary points of $H_{\rm cl}$ in all four variables 
correspond to $\xi,\eta\=0$ and $\phi,j_z$ determined as stationary points of the function $h_{\rm cl}$.
Since the quadratic minimum in variables $\xi,\eta$ has a null index, any stationary point of $H_{\rm cl}$ has an index $r$ equal to that of the corresponding stationary point of $h_{\rm cl}$.
For the determination of \ESQPTs\  it is therefore sufficient to find and classify stationary points of the function $h_{\rm cl}$ on a ball.
Non-degenerate stationary points of $h_{\rm cl}$ with index $r\=0$, $1$, and $2$ cause, respectively, an upward jump, logarithmic divergence and a downward jump of $\partial\varrho_\textnormal{cl}/\partial E$.

Taking into account that the effective coupling parameter $\lambda_{\rm eff}$ in the scaled classical Hamiltonian is reduced with respect to actual $\lambda$ by a factor $\sqrt{2j/N}$, see the text below Eq.\,\eqref{Hcl}, we obtain from Eq.\,\eqref{lc0} the following $j$-dependent values of the critical couplings:
\begin{equation}
\lambda_c(j)=\sqrt{\frac{N}{2j}}\,\frac{\sqrt{\omega\omega_0}}{1+\delta}
\,,\qquad
\lambda_0(j)=\sqrt{\frac{N}{2j}}\,\frac{\sqrt{\omega\omega_0}}{1-\delta}
\label{lamc0scal}
\,.
\end{equation}
These will play the roles of $\lambda_c$ and $\lambda_0$ in individual subspaces of states with fixed values of $j$.
Note that in the highest subspace with $j=N/2$ both expressions in Eq.\,\eqref{lamc0scal} yield the bare values of critical couplings.
A straightforward analysis leads to the following conclusions:

(i)
Stationary points with $r\=0$ represent the global minimum of both $h_{\rm cl}$ and $H_{\rm cl}$ functions, demarcating the ground state of the $N\to\infty$ system.
The minimum appears at $j_z\=-j$ ($\phi$ arbitrary) for $\lambda\<\lambda_c(j)$, and at $j_z=-j{\lambda_c(j)}^2/\lambda^2$, $\phi\=0$ and $\pi$ (a pair of degenerate minima) for $\lambda\geq \lambda_c(j)$.
The ground-state energy is given by the formula
\begin{equation}
E_0(j)=\left\{\begin{array}{ll}
-\omega_0j & {\rm for\ }\lambda\in[0,\lambda_c(j))\,,\\
-\frac{1}{2}\omega_0 j\left[\frac{\lambda_c(j)^2}{\lambda^2} +\frac{\lambda^2}{\lambda_c(j)^2}\right] & {\rm for\ }\lambda\in[\lambda_c(j),\infty)\,,
\end{array}
\right.
\label{E0}
\end{equation}
which exhibits a second-order \QPT\ from normal to superradiant ground-state phase at the critical coupling $\lambda_c(j)$, where $d^2E_0/d\lambda^2$ has a discontinuity.

(ii)
Stationary points with $r\=1$ represent saddles of $h_{\rm cl}$.
They are located at $j_z\=-j$ ($\phi$ arbitrary) for $\lambda_c(j)\leq\lambda\<\lambda_0(j)$, and at $j_z\=-j\lambda_0(j)^2/\lambda^2$, $\phi\=\pi/2$ and $3\pi/2$ for $\lambda\geq\lambda_0(j)$.
These stationary points correspond to an \ESQPT\ (a logarithmic divergence of $\partial\varrho_\textnormal{cl}/\partial E$) at the critical energy
\begin{equation}
E_{c1}(j)=\left\{\begin{array}{ll}
-\omega_0j & {\rm for\ }\lambda\in[\lambda_c(j),\lambda_0(j))\,,\\
-\frac{1}{2}\omega_0 j\left[\frac{\lambda_0(j)^2}{\lambda^2} +\frac{\lambda^2}{\lambda_0(j)^2}\right] & {\rm for\ }\lambda\in[\lambda_0(j),\infty)\,.
\end{array}
\right.
\label{Ec1}
\end{equation}
For $\delta\!\to\!1$ we have $\lambda_0(j)\!\to\!\infty$ and Eq.\,(\ref{Ec1}) is reduced to its first line.

(iii)
Stationary points with $r\=2$ are maxima of $h_{\rm cl}$ at $j_z=-j$ ($\phi$ arbitrary) for $\lambda\geq\lambda_0(j)$ and at $j_z\=+j$ ($\phi$ arbitrary) for $\lambda\geq 0$.
Related \ESQPTs\  (downward jumps of $\partial\varrho_\textnormal{cl}/\partial E$) appear at critical energies
\begin{eqnarray}
E_{c2}(j)=-\omega_0j &&\quad {\rm for\ }\lambda\in[\lambda_0(j),\infty)
\label{Ec2}
\,,\\
E_{c3}(j)=+\omega_0j &&\quad {\rm for\ }\lambda\in[0,\infty)
\label{Ec3}
\,.
\end{eqnarray}
The second maximum of $h_{\rm cl}$ at energy \eqref{Ec3} is the global one, so for $E\geq E_{c3}$ the formula \eqref{rhocl2} yields a constant (saturated) value of the level density equal to $\varrho_{\rm cl}=2j/\omega$.

\begin{figure}[!t]
\includegraphics[width=0.65\linewidth]{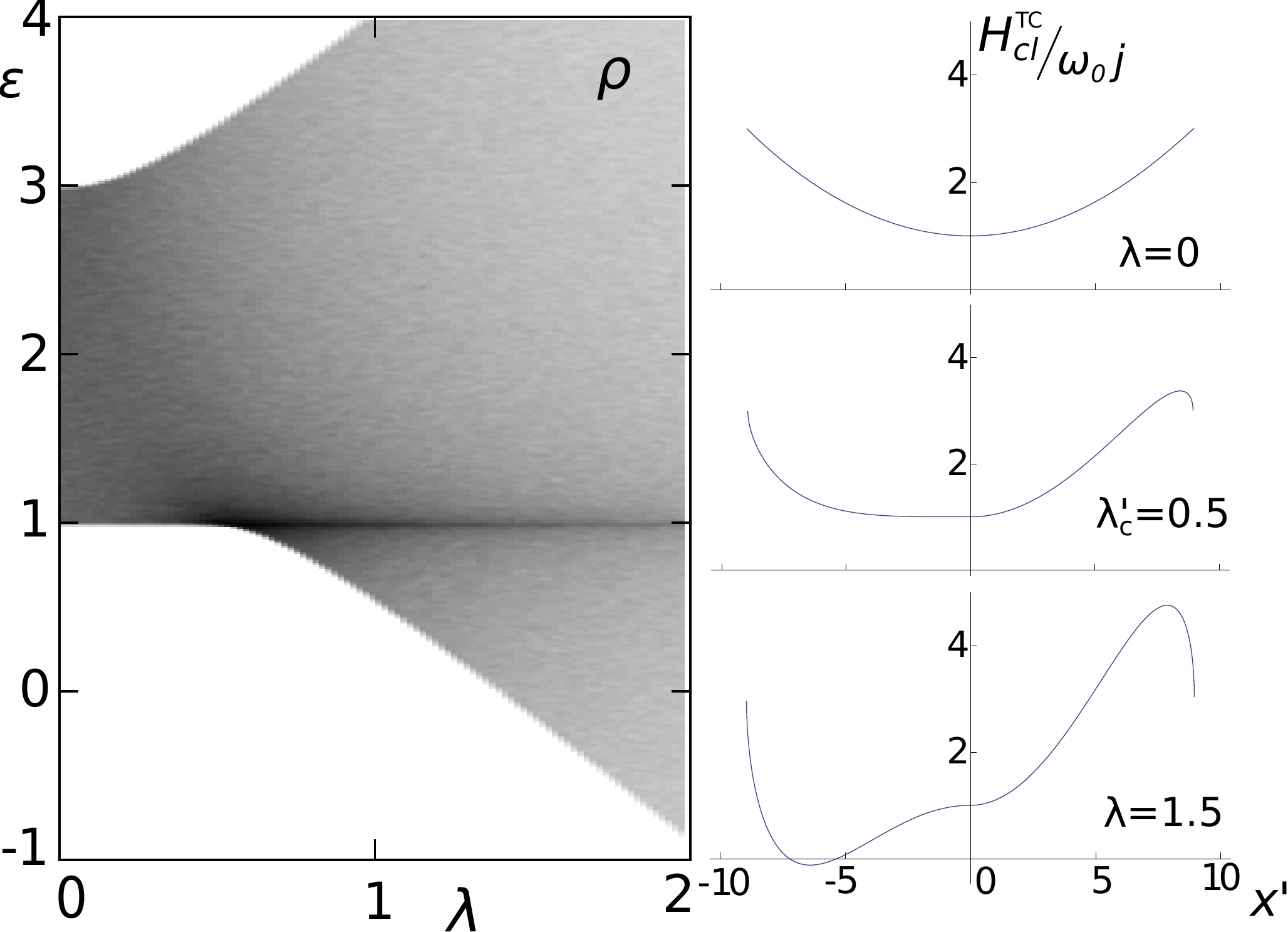}
\caption{
Semiclassical level density in the $M\!=\!2j$ subspace of the $j\!=\!N/2$ Tavis-Cummings model ($\delta\!=\!0$) with $\omega\!=\!2,\omega_0\!=\!1$ as a function of $\lambda$ and $\varepsilon\equiv E/\omega_0j$ (shade plot on the left, with dark areas indicating larger values and vice versa) and three dependences of the corresponding Hamiltonian function \eqref{Hcl1D} on $x'$ for $p'=0$ (on the right).
An \ESQPT\ due to an inflection point of the classical Hamiltonian above the \QPT\ critical point \eqref{lamc1D} results in a logarithmic divergence of the semiclassical level density (the dark band in the shade plot), cf.\,Fig.\,\ref{f_cary}b.}
\label{f_TCdens}
\end{figure}

All \ESQPT\ critical borderlines $E_{c1}$, $E_{c2}$ and $E_{c3}$ from Eqs.\,\eqref{Ec1}--\eqref{Ec3} for various values of $\delta$ are demarcated in Fig.\,\ref{f_qpd} above.
Their existence is numerically verified in Fig.\,\ref{f_dender}, which depicts the $\lambda\times E$ dependence of $\partial\varrho_{\rm cl}/\partial E$ for $j\=N/2$ somewhere in between the Dicke and Tavis-Cummings limits.
The shade plot was obtained through the phase-space integration in Eq.\,\eqref{rhocl2} and a comparison with results obtained by a numerical diagonalization is shown in the two panels on the right. 
We stress that due to the $f\=2$ character of our model, the \ESQPT\ singularities occur in the {\em first derivative\/} of the level density with respect to energy.
However, the conservation of quantity \eqref{M} in the $\delta\=0$ limit and the corresponding reduction of the number of effective degrees of freedom to $f\=1$ (see Sec.\,\ref{INT}) leads to a possibility to generate an \ESQPT\ singularity in the {\em level density itself}.

Such an effect can really be identified within a particular {\em single $M$-subspace\/} of the full model.
The lowest-energy state of any of such subspaces for $N\!\to\!\infty$ coincides with the global minimum of classical Hamiltonian \eqref{Hcl1D} within the available (for a given $M$) domain of the phase space.
It turns out that the most interesting $M$-subspace is the one with $M\=2j$.
This particular subset of states exhibits for $\omega\>\omega_0$ a second-order ground-state \QPT\ at the coupling strength $\lambda$ equal to a critical value
\begin{equation}
\lambda'_{c}(j)=\sqrt{\frac{N}{2j}}\,\frac{\omega\-\omega_0}{2}
\label{lamc1D}
\,.
\end{equation}
At this coupling, the main minimum of function \eqref{Hcl1D} moves from $(x',p')\=(0,0)$ away to $x'\<0$  while $(0,0)$ becomes an inflection point. 
The inflection point is present for $\lambda\in(\lambda'_c(j),\infty)$ and generates an \ESQPT\ (a logarithmic divergence of the level density in the $M\=2j$ subspace) at energy $E_{c3}$ from Eq.\,\eqref{Ec3}.
Note that for $\omega\<\omega_0$ the \ESQPT\ appears for $\lambda\>|\lambda'_c(j)|$ at the upper energy of the unperturbed spectrum.

This effect, addressed already in Ref.\,\cite{Per11a}, is demonstrated in Fig.\,\ref{f_TCdens}.
It displays a shade plot of the semiclassical density of levels with $M\=2j$ (obtained by the $f\=1$ phase--space integration) and the underlying forms of the Hamiltonian \eqref{Hcl1D}.
A finite-size sample of the predicted singularity in the level density was seen in panel (b) of Fig.\,\ref{f_cary} above.
A more detailed analysis of Eq.\,\eqref{Hcl1D} shows that the present type of criticality is absent in the $M\!\neq\!2j$ subspaces. 
This can be intuitively understood from the evaluation of matrix elements $H'_{ii'}$ of the interaction Hamiltonian with $\delta\=0$ in the unperturbed basis, see Eq.\,\eqref{Hamoun}.
The interaction matrix elements quantify the mixing induced by $H_{\rm int}$ in the unperturbed eigenbasis of $H_{\rm free}$.
Neglecting the trivial state with $M\=0$ which does not mix at all, the matrix elements $H'_{ii'}$ are particularly small in the $M\=1$ subspace (dimension 2) and for the $i\=2j$ state of the $M\=2j$ subspace (dimension $2j\+1$).
Only the latter state can develop a singularity in the $j,N\to\infty$ limit.
It is the state $\ket{m\=+j}_{\rm A}\ket{n\=0}_{\rm F}$, which for $\omega\>\omega_0$ represents the lowest state of the $M\=2j$ subspace, while for $\omega\<\omega_0$ it is the highest state; cf.\,panels (b) and (c) of Fig.\,\ref{f_cary}.
Although the $M\!\neq\!2j$ subspaces show no quantum critical effects, it is interesting to realize that a pile up of all $M$-subspaces produces the downward jump of $\partial\varrho_{\rm cl}/\partial E$ (with $\varrho_{\rm cl}$ standing for the total semiclassical level density), as observed in the $E\=E_{c3}$ \ESQPT\ of the $\delta\=0$ system.

\subsection{Quantum phases}
\label{PHA}

The critical borderlines $E_{c1}$, $E_{c2}$ and $E_{c3}$ in the $\lambda\times E$ plane separate spectral domains that we consider to constitute different \lq\lq quantum phases\rq\rq\ of the model (in analogy to thermodynamic phases).
They are denoted by acronyms \D\ (Dicke), \TC\ (Tavis-Cummings), \N\ (normal) and \S\ (saturated), see Fig.\,\ref{f_qpd}.
Both phases \D\ and \TC, which coexist for $\delta$ in between 0 and 1, contain quantum states of a superradiant nature because their energy is lowered with respect to the minimal non-radiant value $E\=-\omega_0j$.
In the limits of $\delta$, the model shows only one type of the superradiant phase: \TC\ for the Tavis-Cummings limit $\delta\=0$ and \D\ for the Dicke limit $\delta\=1$.
In contrast, both phases \N\ and \S\ contain states that resemble excitations in the fully non-radiant regime at $\lambda\=0$.
The \S\ phase above $E_{\rm c3}$ yields a constant, saturated value of level density $\varrho_{\rm cl}$. 

\begin{figure}[!t]
\includegraphics[width=\linewidth]{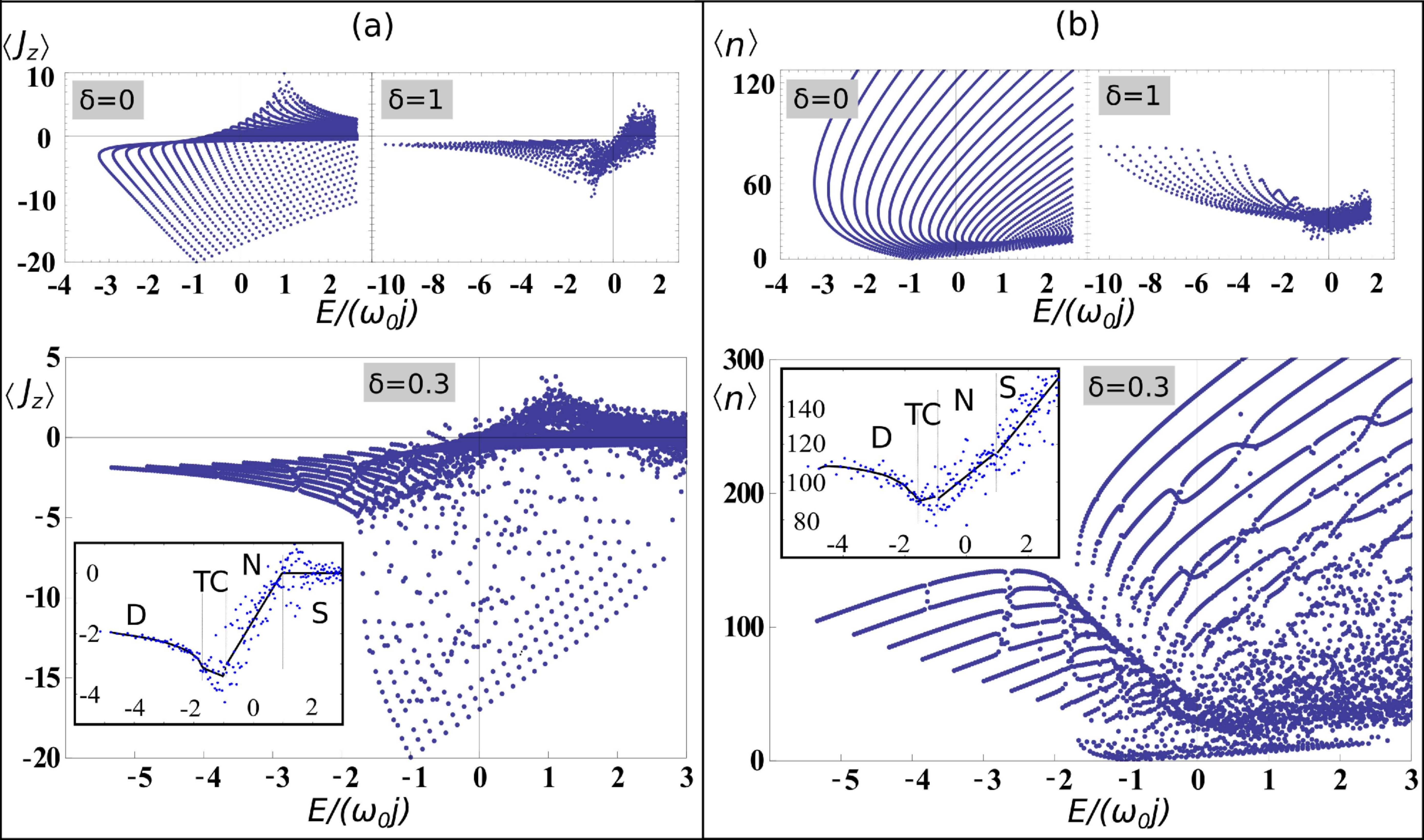}
\caption{Peres lattices of quantum expectation values of observables $J_z$ (panel a) and $n$ (panel b) in individual eigenstates of the $j\!=\!N/2$ Hamiltonian (\ref{H}) for $N\!=\!40$, $\omega\!=\!\omega_0\!=\!1$, $\lambda\!=\!2.5$. The upper panels depict lattices for the limiting values of $\delta\!=\!0$ and 1, whereas the lower panels correspond to $\delta\!=\!0.3$. The insets of the $\delta\!=\!0.3$ panels show smoothed dependences of both observables on energy (averages over 20 neighboring eigenstates and their piecewise fits in various quantum phases).
}
\label{f_Pere}
\end{figure}

While the definition of \ESQPTs\  is obvious from the behavior of quantities like level density at the critical energies, the meaning of quantum phases in between the critical energies is not a priori clear.
It should be looked for in the structure of the energy eigenstates in the corresponding energy domains.
However, sample wave functions in Fig.\,\ref{f_wav} [where panels (a), (b), (d), (f) and (h) correspond to phases \D, \D, \TC, \N\ and \N, respectively] indicate that individual eigenstates do not show sufficient clues for the identification of phases.
We therefore resort to a more efficient visualization tool allowing us a bulk inspection of the eigenstate properties, namely to the method of so-called {\em Peres lattices\/} \cite{Per84}.
It was proven useful in various models (see Ref.\,\cite{Str09} and references therein) including the Dicke model \cite{Bas14,Bas15}.
A general Peres lattice shows expectation values $\ave{P}_i\=\matr{\psi_i}{P}{\psi_i}$ of a selected observable $P$ in individual energy eigenstates $\ket{\psi_i}$ (enumerated by integer $i$) arranged into lattices with energy $E_i$ on one of the axes.
Examples of Peres lattices for observables (a) $J_z\=n^*\-j$ and (b) $n\=b^{\dag}b$ are shown in the respective panels of Fig.\,\ref{f_Pere}.
Their comparison with the map of chaos in Fig.\,\ref{f_qpd}d shows---in agreement with the original conjecture \cite{Per84}---an overall correlation of orderly arranged lattice domains with more regular regions of classical dynamics and vice versa.

The $\delta\=0.3$ Peres lattices in the main panels of Fig.\,\ref{f_Pere} cross all four quantum phases of the model.
There exist apparent similarities between parts of both $\delta\=0.3$ lattices located in the \TC\ and \D\ phases, respectively, and the corresponding Tavis-Cummings and Dicke lattices displayed in the $\delta\=0$ and 1 upper panels.
However, a more specific distinction of quantum phases results from an averaging of both lattices over the neighboring eigenstates, as shown in the insets of both lower panels of Fig.\,\ref{f_Pere}.
Points in these {\em smoothed lattices\/} represent averages $\overline{\ave{J_z}_i}$ and $\overline{\ave{n}_i}$
of the respective expectation values $\ave{J_z}_i$ and $\ave{n}_i$ over 20 neighboring eigenstates.
Smoothed dependences of the averaged quantities on energy are given by lines, resulting from piecewise fits within the four quantum phases.
We observe that various quantum phases are recognized by different characters of the energy dependences, namely:
(i) the phase \D\ by slowly descending dependences of both $\overline{\ave{J_z}_i}$ and $\overline{\ave{n}_i}$ averages, 
(ii) the phase \TC\ by roughly constant dependences,
(iii) the phase \N\ by linearly increasing dependences, and
(iv) the phase \S\ by saturated $\overline{\ave{J_z}_i}\!\approx\!0$ and linearly increasing $\overline{\ave{n}_i}$.

\begin{figure}[!t]
\includegraphics[width=0.55\linewidth]{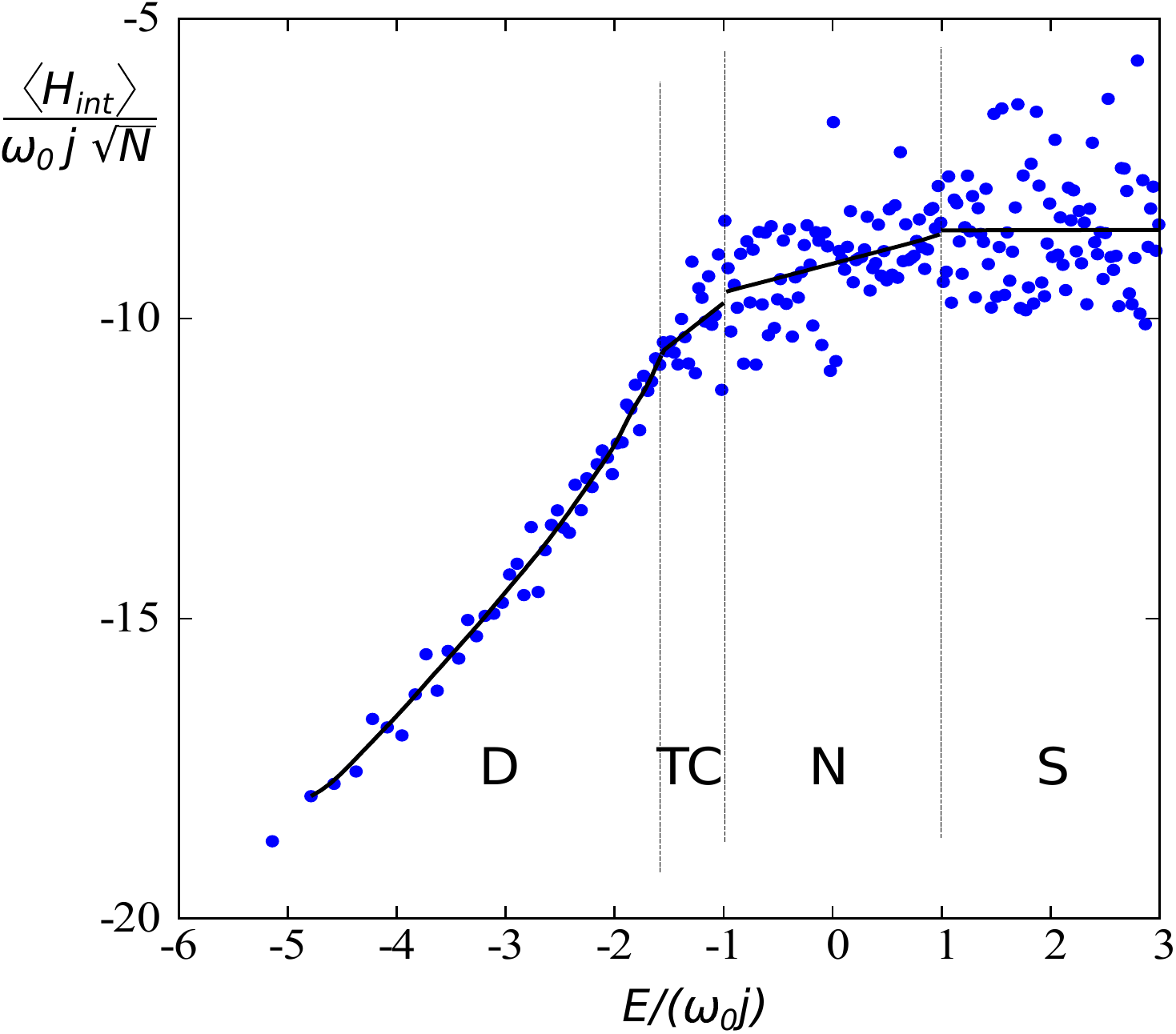}
\caption{Average slopes of the bunches of neighboring 20 levels for the $j\!=\!N/2$ Hamiltonian (\ref{H}) with $N\!=\!40$, $\omega\!=\!\omega_0\!=\!1$ and $\delta\!=\!0.3$ at $\lambda\!=\!2.5$. Quantum phases \D, \TC, \N\ and \S\ are distinguished by different smoothed energy dependences of the average slopes---see the piecewise fits indicated by full black lines.}
\label{f_slo}
\end{figure}

Peres lattices for both quantities $J_z$ and $n$ are connected with the lattice of the interaction Hamiltonian $H_{\rm int}$ through the energy conservation.
Using the Hellmann-Feynman formula, $dE_i/d\lambda\=\ave{dH/d\lambda}_i$, we derive the following relation for the slope of individual energy levels
\begin{equation}
\frac{dE_i}{d\lambda}=\frac{\ave{H_{\rm int}}_i}{\sqrt{N}}
=\frac{E_i-\omega_0\ave{J_z}_i-\omega\ave{n}_i}{\lambda}
\label{slope}
\,.
\end{equation}
The slopes averaged over bunches of neighboring 20 levels, as in the insets of Fig.\,\ref{f_Pere}, are presented in Fig.\,\ref{f_slo} for the same values of control parameters. 
Piecewise fits of the energy dependences in individual quantum phases are again indicated by lines.
We see that any of the \D, \TC, \N\ and \S\ quantum phases is characterized by a specific, roughly invariant energy dependence of the averaged level slope $\overline{dE_i/d\lambda}$ within the corresponding energy domain.
Abrupt changes of the character of these dependences coincide with critical \ESQPT\ energies.
This is in accord with a general relation between the density and flow signatures of \ESQPTs, namely with the fact that a typical \ESQPT\ generates the same type of non-analyticity in the energy dependences of both quantities $\varrho_{\rm cl}$ and $\overline{dE_i/d\lambda}$ \cite{Str14,Str16}.
Indeed, the discontinuities of $\partial(\overline{dE_i/d\lambda})/\partial E$ at the critical energies $E_{c2}$ and $E_{c3}$, as observed in Fig.\,\ref{f_slo}, are consistent with the analogous behavior of the $\partial\varrho_{\rm cl}/\partial E$ (Fig.\,\ref{f_dender}).
On the other hand, the anticipated point of a singular growth of the level slope (logarithmic divergence of its energy derivative) at $E_{c1}$ is smoothed out in the finite spectrum for a moderate system's size.

\section{Atom--field and atom--atom entanglement}
\label{ENT}
\subsection{Measures of bipartite entanglement}
\label{EME}

We turn to the study of quantum entanglement properties of individual eigenstate of the model Hamiltonian and their links to the \ESQPTs.
Our analysis includes two types of bipartite entanglement: (a) that between the bosonic field and the set of all atoms (atom--field entanglement), and (b) that between any pair of individual atoms (atom--atom entanglement).
The entanglement of type (a) is an important ingredient of superradiance since the interaction term of the Hamiltonian \eqref{H} carries a direct coupling between the atomic and field subsystems.
In contrast, the entanglement of type (b) appears only due to an indirect coupling of individual atoms via the bosonic field, so it may be expected to be just a \lq\lq higher-order\rq\rq\ effect. 
We start with a brief description of the measures used to quantify both types of entanglement.

In the following, the atom--field and atom--atom entanglement will be evaluated in individual eigenstates $\ket{\psi_i}\in{\cal H}^{j,l}_{\rm A}\otimes{\cal H}_{\rm F}$ of Hamiltonian \eqref{H}.
Some examples of the eigenstate wave functions were shown in Fig.\,\ref{f_wav}.
Since the wave-function arguments $\phi$ and $x$, respectively, correspond directly to the atomic and field coordinates, a compound state $\ket{\psi}$ is factorized with respect to the atom--field partitioning of the system if it has a product wave function $\psi(\phi,x)=\psi'_{\rm A}(\phi)\psi''_{\rm F}(x)$, where $\psi'_{\rm A}$ and $\psi''_{\rm F}$ are arbitrary atomic and field wave functions.
The method to quantify a departure of a given pure state $\ket{\psi}$ from exact factorization, i.e., an amount of atom--field entanglement involved in $\ket{\psi}$, makes use of the {\em von Neumann entropy\/} corresponding to the reduced density operators $\rho_{\rm A}$ and $\rho_{\rm F}$ of the atomic and field subsystem, respectively \cite{Ved02}.
These operators are obtained by partial tracing of the total density operator $\rho=\ket{\psi}\bra{\psi}$ over the irrelevant part of the total Hilbert space, that is $\rho_{\rm A}={\rm Tr}_{\rm F}\rho$ (trace goes over $\mathcal{H}_{\rm F}$) for the atomic subsystem and $\rho_{\rm F}={\rm Tr}_{\rm A}\rho$ (trace goes over $\mathcal{H}^{j,l}_{\rm A}$) for the field subsystem.
Von Neumann entropy of the pure compound state $\rho$ is by definition zero, but entropies of both reduced density operators $\rho_{\rm A}$ and $\rho_{\rm F}$ satisfy: $S_{\rm A}=S_{\rm F}\geq 0$.
The case $S_{\rm A}\=S_{\rm F}\=0$ implies a separable compound state, while the case $S_{\rm A}\=S_{\rm F}\>0$ indicates that the reduction of $\rho$ to a single subsystem leads to a loss of information on mutual entanglement between subsystems.
A maximally entangled state yields $S_{\rm A}\=S_{\rm F}\=\ln(2j\+1)$, where $2j\+1$ is a dimension of $\mathcal{H}_{\rm A}^{j,l}$, the smaller of both subspaces.
We therefore define a normalized atom--field entanglement entropy in a compound state $\ket{\psi}$ as
\begin{equation}
S(\psi)=-\frac{\textnormal{Tr}\left[\rho_{\textnormal{A}}\,\textnormal{ln}\rho_{\textnormal{A}}\right]}{\ln(2j\+1)}
=-\frac{\textnormal{Tr}\left[\rho_{\textnormal{F}}\,\textnormal{ln}\rho_{\textnormal{F}}\right]}{\ln(2j\+1)}
\label{S}
\,.
\end{equation}
It changes between $S\=0$ for separable states and $S\=1$ for maximally entangled states.

Quantifying the atom--atom entanglement, i.e., quantum correlations between a randomly chosen pair $\{k,l\}$ of atoms, is a more complex problem.
The use of the above entropic approach is disabled by the fact that for any state $\ket{\psi}$ of the whole atom--field system, an arbitrary pair of atoms generically occurs in a {\em mixed\/} quantum state. 
It is known that mutual entanglement of a pair of objects in a mixed compound state cannot be recognized by a non-zero entropy of the reduced density operators \cite{Per96}.
Indeed, a mixed compound state of the pair generates mixed reduced states of the objects even in absence of entanglement.
In such cases, the evaluation of an entropy-based measure of entanglement (so-called entanglement of formation) has to be performed with respect to all possible decompositions of the compound density operator into statistical mixtures of pure states \cite{Ben96}, which is from the computational viewpoint a difficult task \cite{Hua14}.

A way to bypass this obstacle for two-qubit systems was proposed in Refs.\,\cite{Hil97, Woo98} in terms of a quantity called {\em concurrence}.
The idea was applied and further elaborated \cite{Ost02,Vid04a,Vid04b,Wan02,Wan03} for multi-qubit systems in fully symmetric states , like our ensemble of $N$ two-level atoms with $j=N/2$.
The atom--atom entanglement in this case is characterized by a scaled concurrence
\begin{equation}
C(\psi)=(N\-1)\,\textnormal{max}
\left\{\sqrt{\lambda_1}\-\sqrt{\lambda_2}\-\sqrt{\lambda_3}\-\sqrt{\lambda_4}\,,0\right\}
\label{C}
\,,
\end{equation}
where $\lambda_1\geq\lambda_2\geq\lambda_3\geq\lambda_4$ are eigenvalues (real and non-negative) of a non-Hermitian matrix $\rho_{\rm A}^{kl}\,\tilde{\rho}_{\rm A}^{kl}$, with $\rho_{\rm A}^{kl}$ denoting a reduced density matrix of the pair of atoms $\{k,l\}$ and $\tilde{\rho}_{\rm A}^{kl}\equiv\left(\sigma_y^k\otimes\sigma_y^l\right)\rho_{\rm A}^{kl*}\left(\sigma_y^k\otimes\sigma_y^l\right)$ the corresponding \lq\lq spin-flipped\rq\rq\ (conjugate under the time reversal) state (with $\otimes$ denoting the tensor product of operators for the selected atoms and $\rho_{\rm A}^{kl*}$ a complex conjugated density matrix in the $\sigma_z^k,\sigma_z^l$ eigenbasis).
The argument $\psi$ in the definition \eqref{C} again reminds the compound state $\ket{\psi}$ in which the concurrence is calculated.
It has to be stressed that the full symmetry of $\ket{\psi}$ under the exchange of atoms, which is guaranteed in the $j=N/2$ subspace, ensures that the reduced density matrix $\rho_{\rm A}^{kl}$, as well as $\tilde{\rho}_{\rm A}^{kl}$ and $C$, are the same for {\em any\/} pair of atoms, hence do not in fact depend on $k,l$.

\begin{figure}[t!]
\includegraphics[width=0.75\linewidth]{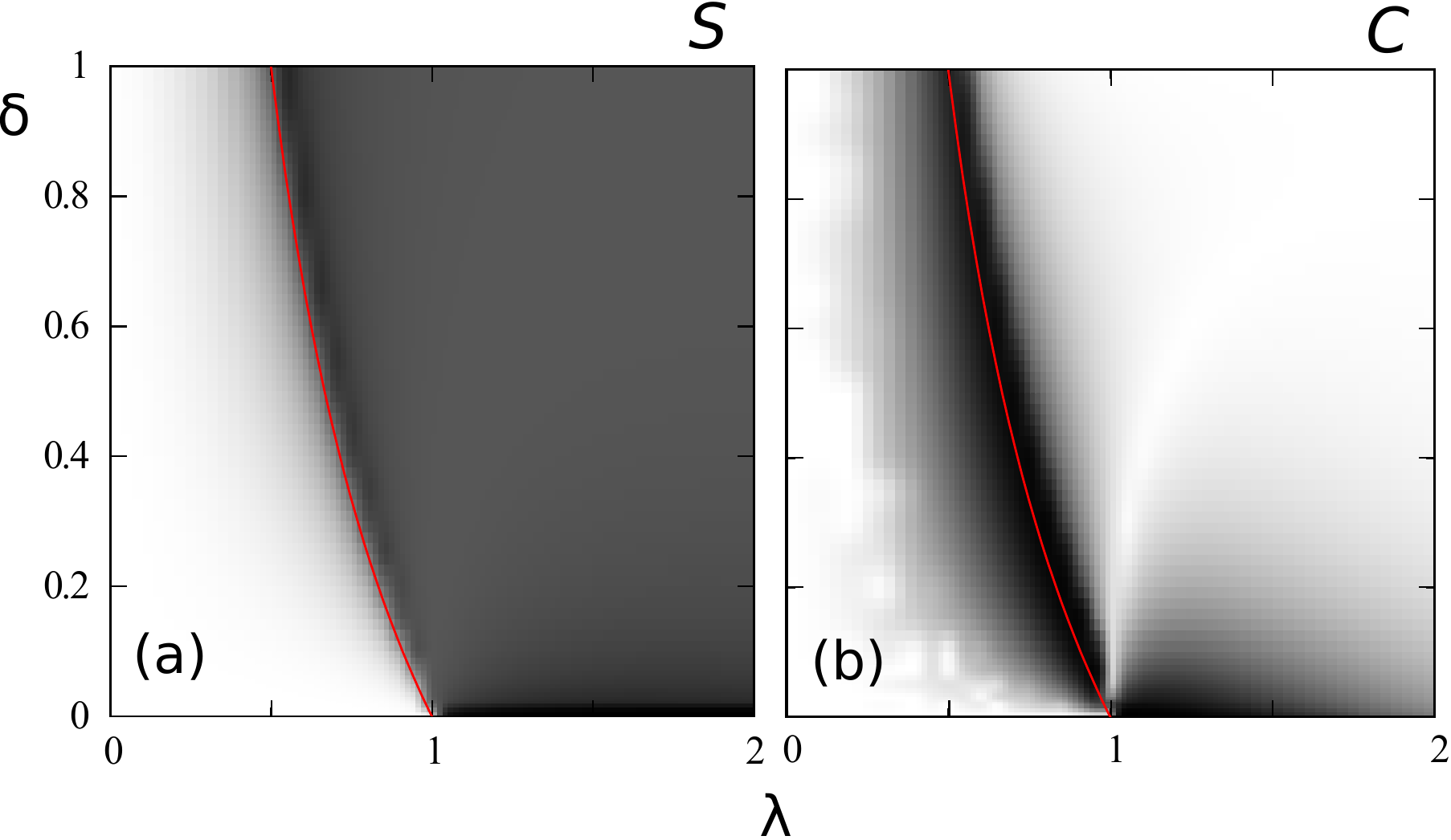}
\caption{(Color online) Entanglement properties of the ground state of the $j=N/2$ model in the plane of control parameters $\lambda$ and $\delta$ (for $\omega\!=\!\omega_0\!=\!1$, $N=40$). The red curve indicates the \QPT\ critical coupling $\lambda_c$. Left: atom--field entanglement measured by the entropy \eqref{S}. Right: atom--atom entanglement measured by the concurrence \eqref{C}. The darker areas indicated larger entanglement.}
\label{f_eng}
\end{figure}

Matrix elements of the reduced two-atom density matrix $\rho_{\rm A}^{kl}$ for any symmetric state of an $N$-qubit system can be expressed through expectation values $\langle J_z\rangle$, $\langle J_z^2\rangle$ and $\langle J_+\rangle$ of collective quasi-spin operators in a given state---see formulas (4) and (11) in Ref.\,\cite{Wan02}.
This makes it possible to perform a straightforward numerical computation of $C$ in the $j=N/2$ subspace of our model.
Moreover, as shown in Ref.\,\cite{Wan03}, the scaled concurrence in this subspace is related to the so-called Kitagawa--Ueda spin-squeezing parameter $\xi^2$ though the relation $C={\rm max}\left\{1-\xi^2,0\right\}$, so it
varies between $C\=0$ for separable states and $C\=1$ for the states that exhibit the maximal entanglement allowed for a given $N$.
An entropy corresponding to the entanglement of formation of a single atomic pair reads as \cite{Woo98}
\begin{equation}
s(\psi)=-\sum_{\sigma=\pm}a_{\sigma}\ln a_{\sigma}
\,,\quad
a_{\pm}=\tfrac{1}{2}\left[1\pm\sqrt{1\-\tfrac{C(\psi)^2}{(N-1)^2}}\right]
\label{saa}
\,.
\end{equation}
For $C\=0$ we have $s\=0$, while for $C\=1$ we obtain $s\approx\ln 2/(N\-1)^2\+\ln(N\-1)/2(N\-1)^2$, which decreases from $s\=\ln 2$ for $N\=2$ to zero in the classical limit $N\!\to\!\infty$.
With $N\gg 1$, the maximal entropy behaves as $s\sim\ln N/N^2$.
Note that the $N\gg 1$ maximal atom--field entanglement entropy {\em per atom\/} decreases as $S/N\sim \ln N/N$, see Eq.\,\eqref{S}, so it exceeds the atom--atom entanglement entropy about $N$ times.
These scaling properties are verified theoretically and by large-$N$ numerical calculations \cite{Lam04}.

The atom--field and atom--atom entanglement properties have been studied for the ground state of the Dicke model \cite{Lam04,Lam05,Bak12}.
As in other models of similar nature, see e.g. Refs.\,\cite{Ost02,Vid04a,Bar06,Vid07}, the second-order \QPT\ was shown to induce a singularity in both entanglement measures \eqref{S} and \eqref{C}.
Figure~\ref{f_eng} verifies this conclusion in our extended model with variable parameter $\delta$.
We observe that both $S$ and $C$ measures exhibit an increase at about the critical coupling $\lambda_c$ from Eq.\,\eqref{lamc0scal}.
For $\delta\>0$, the atom--atom entanglement drops back to nearly zero values with $\lambda\>\lambda_c$.
The atom--field entanglement saturates at a value $S\approx\ln 2/\ln(N\+1)$ for $\lambda\gg\lambda_c$, which is due to an irreducible atom--field coupling in the lowest positive parity state in the strong coupling limit \cite{Lam04}.
In the $\delta\!\approx\!0$ region, close to the integrable Tavis-Cummings limit, both entanglement measures show roughly a step-like increase at $\lambda_c$.
This is due to a specific mechanism, in which the ground state at each $\lambda$ is formed via unavoided crossings of levels with different values of the conserved quantum number $M$, see Fig.\,\ref{f_spe}a.
We will analyze the integrable case in more detail below.

\subsection{Atom--field entanglement: the $\delta\=0$ case}
\label{AFE0}

In the following, we present results of a numerical study of the atom--field entanglement in individual eigenstates of Hamiltonian \eqref{H} with $j\=N/2$.
We start by analyzing the integrable Tavis-Cummings limit $\delta\=0$. 
This simple setting will allow us to obtain some insight into the entanglement properties from quasi-analytic solutions, which will serve as a useful reference for the less trivial $\delta\>0$ case.
We know that for $\delta\=0$ the Hilbert space splits into the subspaces ${\cal H}_M$ with fixed values of quantum number $M$, see Eq.\,\eqref{HilM}.
The reduced density matrix of the atomic subsystem within each ${\cal H}_M$ reads as
\begin{eqnarray}
\rho_\textnormal{\rm A}&=&
\sum_{n=0}^{M}\sum_{m,m'=-j}^{{\rm min}\{M-j,j\}}\!\!\!
\alpha_{mn_M(m)}\alpha^*_{m'n_M(m')}
\ket{m}_{\rm A}\scal{n}{n_M(m)}_{\rm F}\scal{n_M(m')}{n}_{\rm F}\bra{m'}_{\rm A}
\nonumber\\
&=&\sum_{m=-j}^{{\rm min}\{M-j,j\}}\!\!\!|\alpha_{mn_M(m)}|^2\ket{m}_{\rm A}\bra{m}_{\rm A}
\,,
\label{wfENT}
\end{eqnarray}
where $n_M(m)=M\-j\-m$ is a number of bosons associated with quasi-spin projection $m$ for a given value of $M$.
This expression implies that for $\delta\=0$ the entanglement entropy \eqref{S} of any eigenstate is equal to the {\em wave-function entropy\/} \eqref{wentro} corresponding to its expansion in the non-interacting basis.

\begin{figure}[t!]
\includegraphics[scale=0.8,angle=0]{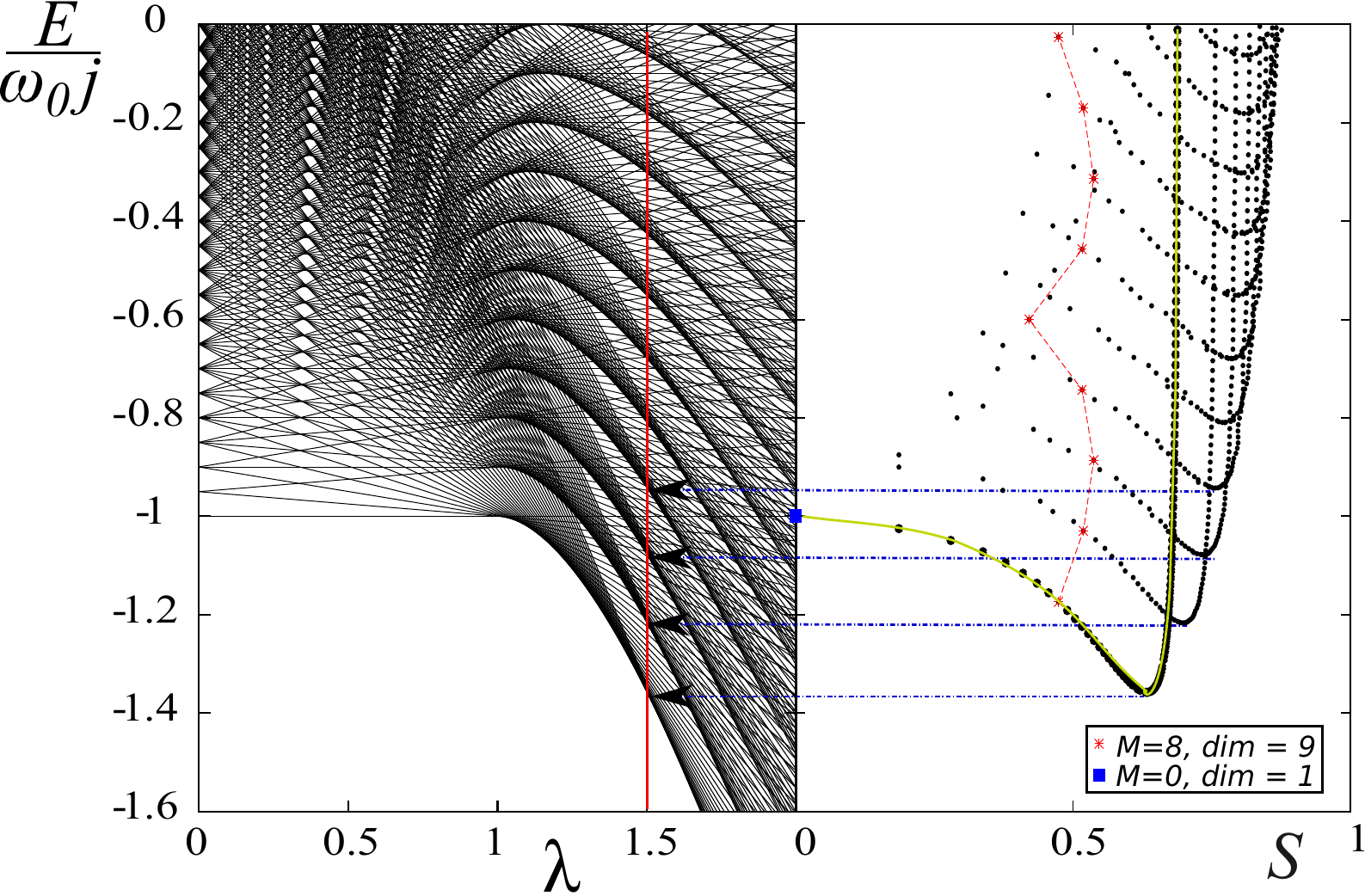}
\caption{(Color online) The full energy spectrum of the $\delta\!=\!0$ model with $\omega\!=\!\omega_0\!=\!1$ for $j\!=\!N/2$ and $N\!=\!40$ (left panel), and the atom--field entanglement entropies $S$ in individual eigenstates corresponding to $\lambda\!=\!1.5$ cut of the spectrum (right panel). The lowest V-shaped chain of points (connected by the green curve) in the entropic spectrum corresponds to the lowest states from various $M$-subspaces, the second chain to the second states etc. The points corresponding to two selected $M$-subspaces ($M\!=\!0$ and 8) are highlighted.
}
\label{f_pileup}
\end{figure}

Let us focus first on the {\em tuned\/} case $\omega\=\omega_0$.
The wave-function entropies for the $M\=2j$ spectrum of a tuned $\delta\=0$ Hamiltonian were shown in Fig.\,\ref{f_cary}a.
We notice that the entropy of individual eigenstates does not change with $\lambda$ and that it has an apparent symmetry with respect to the vertical reflection of the spectrum around the centroid energy $E(0)$.
The former feature was explained by constancy of eigenstates of the simple Hamiltonian \eqref{Hamoun}, the latter follows from a recursive relation for the eigenstate components which yields the same distributions $|\alpha_{mn}|^2$, hence the same entropies, for the pair of levels with opposite slopes.

The full spectrum of the Tavis-Cummings model is obtained by combining the spectra for all values of $M$. 
This is for $\omega\=\omega_0$ seen in Fig.\,\ref{f_pileup}.
It shows the energy spectrum as a function of $\lambda$ (in the left panel) and the atom--field entanglement entropy of individual eigenstates for one particular value of the coupling strength (the right panel).
The \QPT\ critical coupling $\lambda_c$ coincides with the point where the energy of the $M\=0$ ground state is crossed by the lower state from the $M\=1$ space.
A further increase of $\lambda$ above $\lambda_c$ leads to a sequence of consecutive level crossings in which the lowest states from subspaces with increasing $M$ become instantaneous ground states of the system.
A similar mechanism applies also in the spectrum of excited states, where we observe a sequence of separated caustic structures formed by states with increasing ordinal numbers within each $M$-subspace (the $n$th caustic structure in the vertical direction represents an envelope of lines corresponding to the $n$th states from individual $M$-subspaces).

The entropic spectrum on the right of Fig.\,\ref{f_pileup} arises in a similar way.
It is composed of several mutually shifted V-shaped chains of points, each of them containing a collection of states from different $M$-subspaces. 
For example, the lowest V-chain of points is formed by the lowest-energy states from all $M$-subspaces, the second lowest chain by second lowest-energy states and so on.
To indicate a contribution of a single $M$-subspace to the entropic spectrum, we highlighted the points corresponding to two values of $M$.
For $M\>0$, the contribution has a seagull-like form (see the $M\=8$ example in the figure), whose reflection symmetry around $E(0)$ results from the above-discussed properties of Eq.\,\eqref{Hamoun}.
The instantaneous ground state at the given $\lambda\=1.5$, which is well above the critical point $\lambda_c$, has a high value of $M$ and therefore carries a considerable amount atom--field entanglement (as in general expected in superradiance).
On the other hand, the trivial one-dimensional subspace with $M\=0$ formed by the state $\ket{m\=-j}_{\rm A}\ket{n\=0}_{\rm F}$, which was the ground state of the $\delta\=0$ Hamiltonian at $\lambda\<\lambda_c$ and becomes an excited state with energy $E\=E_{c2}\=-\omega_0 j$ for $\lambda\>\lambda_c$, has apparently zero atom--field entanglement.
We will see below that some remnants of this state remain present in entropic spectra also in the non-integrable $\delta\>0$ regime.

\begin{figure}[t!]
\includegraphics[width=\linewidth]{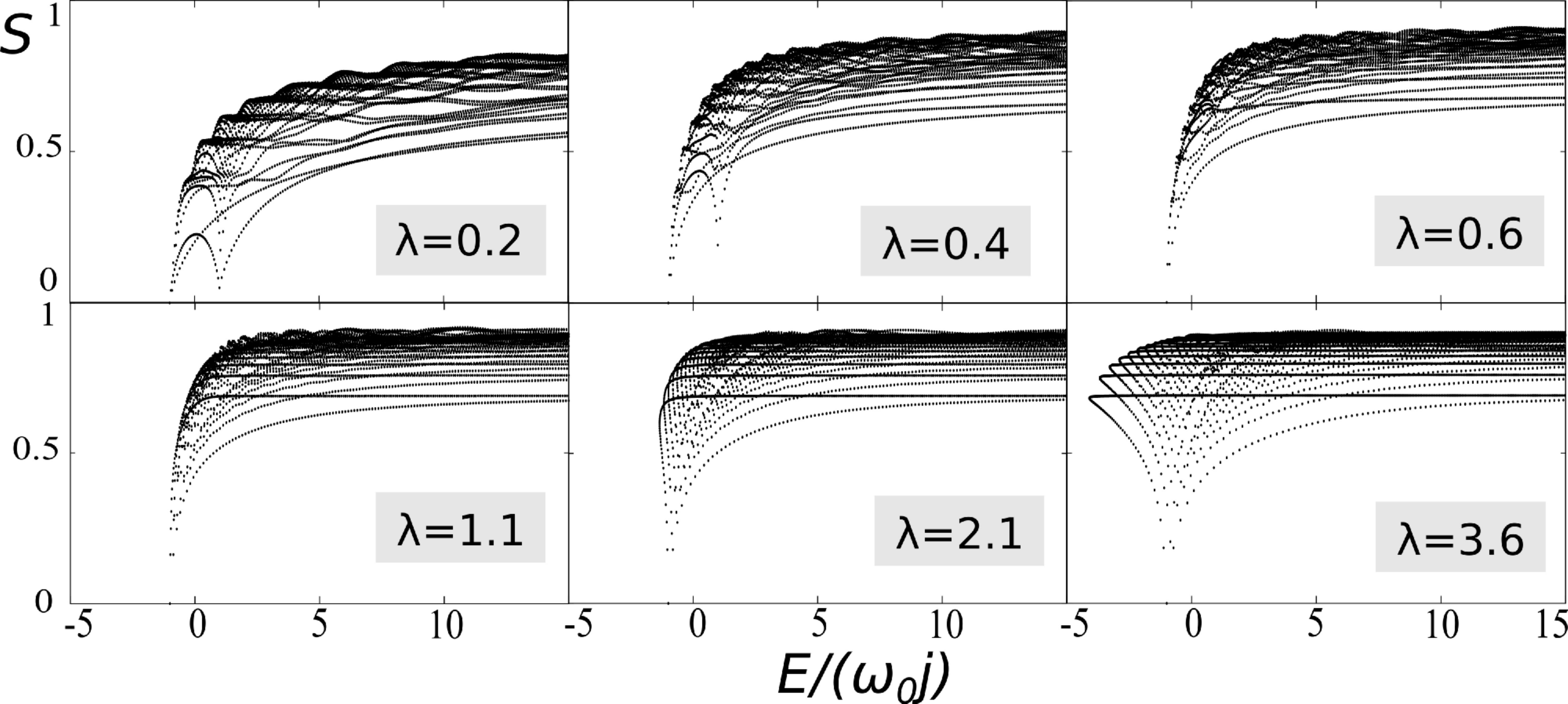}
\caption{The atom--field entanglement entropy of individual eigenstates of the $\delta\!=\!0$ model
with $j\!=\!N/2$ and $N\!=\!40$ in the detuned case ($\omega\!=\!2,\omega_0\!=\!1$) for several values of the coupling strength $\lambda$.}
\label{f_detun}
\end{figure}

Let us consider now the {\em detuned\/} case $\omega\!\neq\!\omega_0$. 
The atom--field entropic spectra of a Hamiltonian with $\delta\=0$ and $\omega\=2\omega_0$ are displayed in Fig.\,\ref{f_detun} for increasing values of the coupling strength $\lambda$ (the critical coupling is $\lambda_c\=\sqrt{2}$).
We know (see Sec.\,\ref{INT}) that in the strong coupling limit, $\lambda\!\gg\!\lambda_c$, the detuned situation becomes very similar to the tuned one.
This is obvious from a comparison of the $\lambda\=3.6$ entropic spectrum in Fig.\,\ref{f_detun} with that in Fig.\,\ref{f_pileup}.
On the other hand, for small coupling strengths, $\lambda\!\lesssim\!\lambda_c$, the detuning $\Delta\omega$ plays an important role.
A detuned system shows generally lower values of the atom--field entanglement than the tuned one, which can be attributed to a smaller perturbation efficiency of the interaction Hamiltonian due to larger energy gaps between the unperturbed (factorized) states, see Eq.\,\eqref{Hamoun}.

It is clear that the factorized state $\ket{m\=-j}_{\rm A}\ket{n\=0}_{\rm F}$ with $M\=0$ at critical energy $E_{c2}\=-\omega_0 j$ is an invariant eigenstate of any $\delta\=0$ Hamiltonian, irrespective of $\omega,\omega_0$ and $\lambda$. 
So the $S\=0$ point at the place of the unperturbed ground state is present in all entropic spectra of Fig.\,\ref{f_detun}.
However, for an $\omega\!\neq\!\omega_0$ system, a local decrease (converging to $S\=0$ with $j,N\to\infty$) of the atom--field entanglement is present also at the highest critical energy $E_{c3}\=+\omega_0 j$.
This is due to quantum critical properties of the $M\=2j$ subspace discussed in Sec.\,\ref{QPT}.
We consider here the $\omega\>\omega_0$ case (see Fig.\,\ref{f_cary}b), but the same phenomenon takes place also in the reversed detuning hierarchy $\omega\<\omega_0$ (Fig.\,\ref{f_cary}c).
For $\lambda\<\lambda'_c$, where $\lambda'_c$ is the critical coupling \eqref{lamc1D}, the lowest state of the $M\=2j$ subspace at energy $E_{c3}$ keeps the factorized form $\ket{m\=+j}_{\rm A}\ket{n\=0}_{\rm F}$ (mind the sign difference in $m$ from the other factorized state, which means that now a maximal possible, for a given $j$, number of atoms is in the {\em upper\/} level).
Distinct lowering of the atom--field entanglement entropy at $E\=E_{c3}$  is observed in the $\lambda\=0.2$ and 0.4 panels of Fig.\,\ref{f_detun} (in which $\lambda'_c\=0.5$).
This \QPT-related effect becomes sharper with increasing $j$ and $N$, and in the $j,N\to\infty$ limit it results in a similar $S\=0$ cusp as that at energy $E_{c2}$.
For $\omega\<\omega_0$, the above factorized state, causing the same effect, is the highest state of the $M\=2j$ subspace at $\lambda\=0$.

\begin{figure}[t!]
\includegraphics[width=0.8\linewidth]{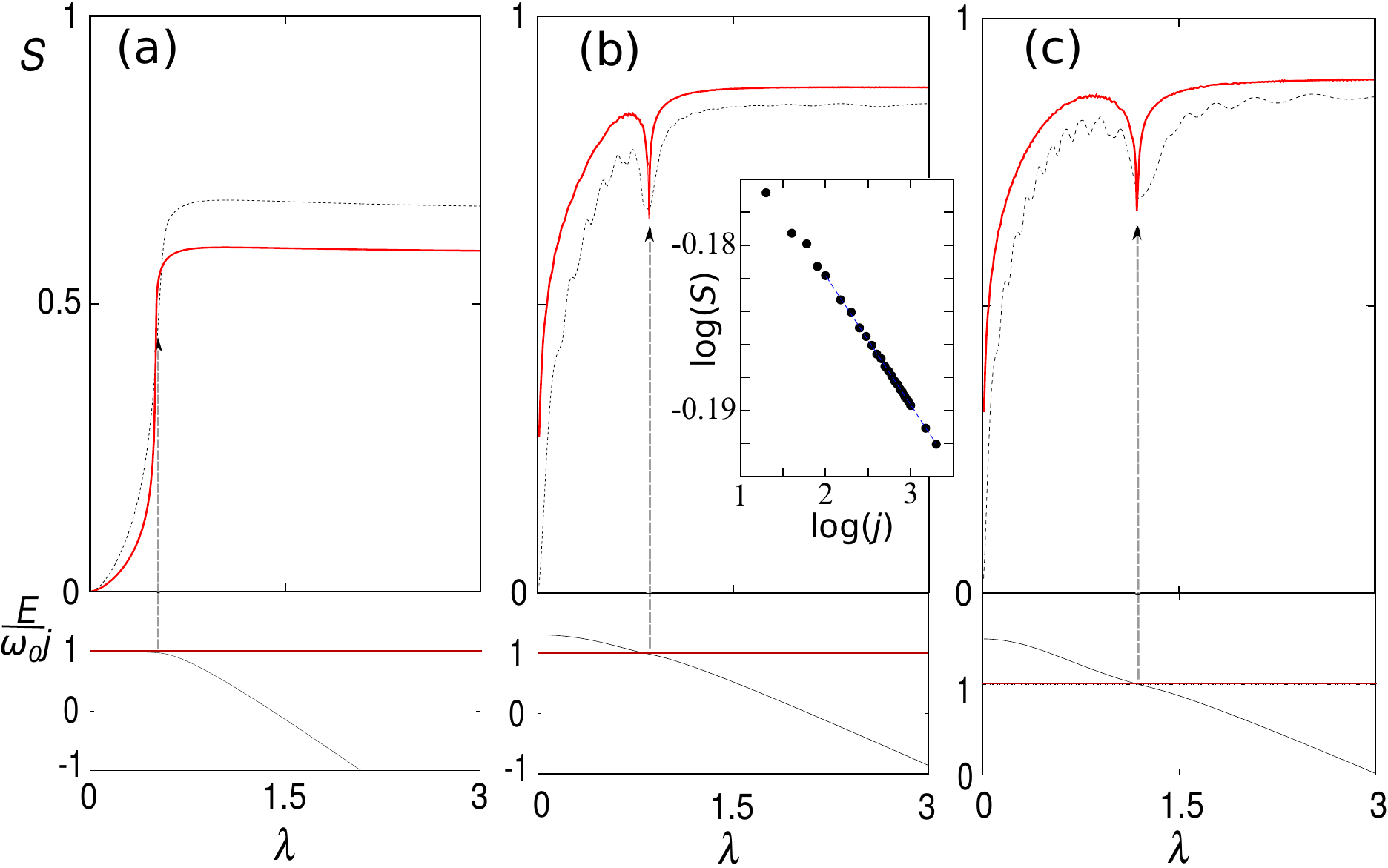}  
\caption{
The atom--field entanglement entropy for three states from the $M\!=\!2j$ subspace of the $\delta\!=\!0$ detuned ($\omega\!=\!2,\omega_0\!=\!1$) model with $j\!=\!N/2$: the ground state (panel a) and excited states which at $\lambda\!=\!0$ have 15\,\% (panel~b) and 24\,\% (panel~c) of the maximal excitation energy. The dashed curves are for $N\!=\!40$, the full ones for $N\!=\!1000$. The energy of the corresponding level (for $N\!=\!40$) is shown at the bottom of each panel. The log-log plot in panel (b) inset shows the minimum value of $S$ in the dip as a function of $j$ for the level at 15\,\% of the spectrum (for this level, the dependence is roughly $S_{\rm min}\propto 1/j^{0.008}$).}
\label{f_entroTC}
\end{figure}

An analogous lowering of the atom--field entanglement entropy at the upper critical energy $E_{c3}$ appears also for the coupling strengths {\em above\/} the \QPT\ critical value $\lambda'_c$.
The effect in this domain is caused by the $f\=1$ type of \ESQPT\ in the $M\=2j$ subspace.
Indeed, when excited states within this subspace cross the critical energy, they get temporarily localized in the coordinate region around the point $x'\=0$, i.e., at the position of the factorized state.
This is due to \lq\lq dwelling\rq\rq\ of the $E\=E_{c3}$ classical trajectory and a semiclassical wave function at an inflection point of the classical Hamiltonian \eqref{Hcl1D}.
An analogous phenomenon was observed also in other quasi $f\=1$ models, see e.g. Ref.\,\cite{Hei06}.
The lowering of the eigenstate entropy in the $E\!\approx\!E_{c3}$ domain was visible already in panels (b) and (c) of Fig.\,\ref{f_cary}, but it is not apparent in Fig.\,\ref{f_detun}.
To see the effect in the pile up of all-$M$ spectra, one would have to go to higher $j,N$ values, when the localization becomes stronger and the entropy decreases to its lower limit $S\=0$.

\begin{figure}[t!]
\includegraphics[width=0.82\linewidth]{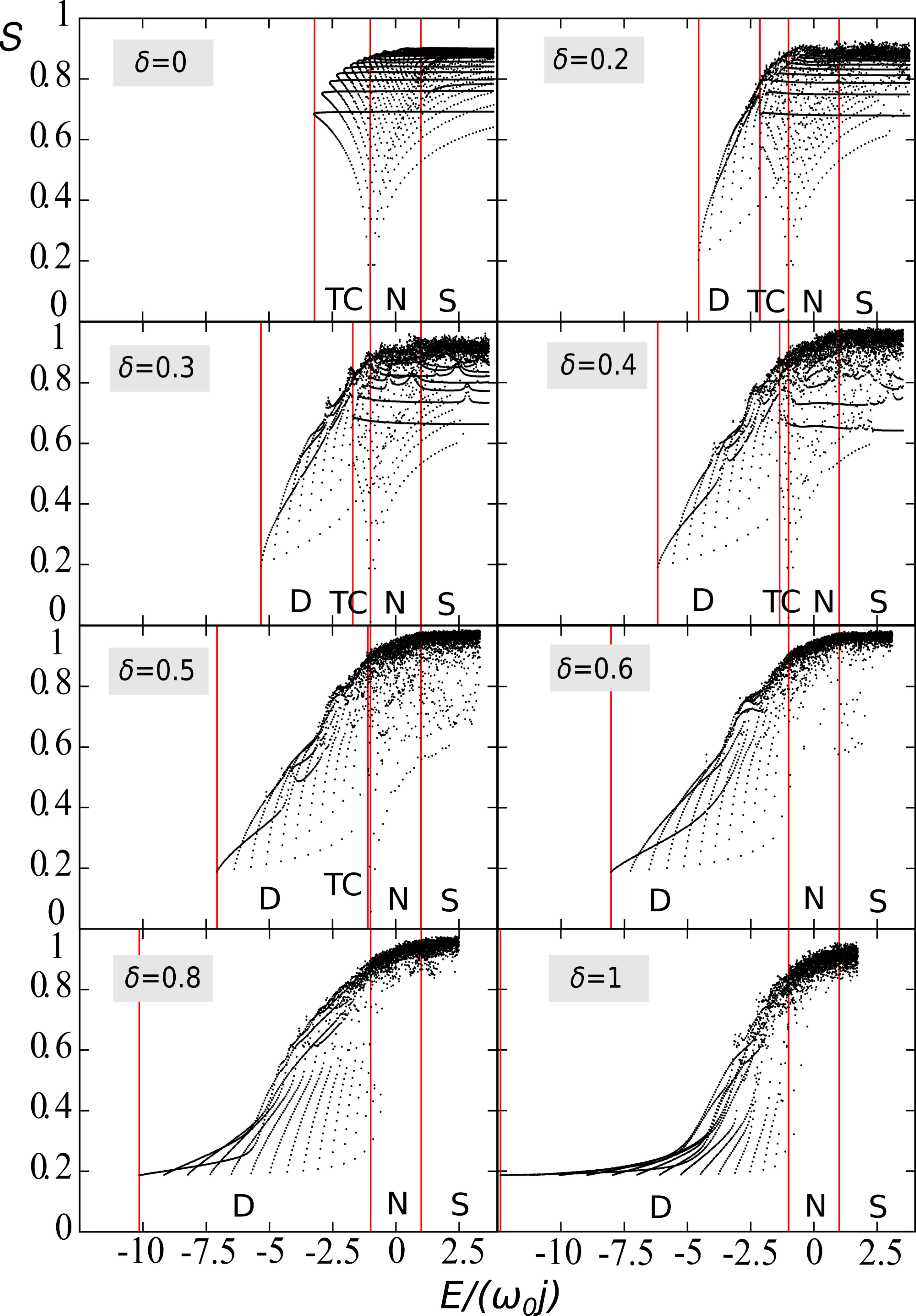}
\caption{Evolution of the $j\!=\!N/2$ entropic spectra of the atom--field entanglement with increasing parameter $\delta\!\in\![0,1]$ at $\lambda\!=\!2.5$, $\omega\!=\!\omega_0$, $N\!=\!40$. The \ESQPT\ energies are indicated by vertical lines. The spectrum contains $5000$ well converged eigenstates.}
\label{f_Sdelta}
\end{figure}

The \ESQPT-based localization effect is nevertheless well documented in Fig.\,\ref{f_entroTC}, which depicts the evolution of the atom--field entanglement for three selected levels (ground state and two excited states) from the $M\=2j\=N$ subspace for $N\=40$ and 1000.
We see that the passage of the selected excited states through the \ESQPT\ energy $E_{c3}$ is correlated with a local decrease of the entanglement entropy [sharp dips in the dependences in panels (b) and (c)].
As demonstrated by the curves for two values of $N$, the drop of entropy becomes sharper and deeper with increasing size of the system.
The decrease of $S$ at the lowest point of the dip with $j\=N/2$ for one of the excited states is shown in the inset of Fig.\,\ref{f_entroTC}b.
The depicted log-log plot indicates a very slow power-low decrease according to $S_{\rm min}\propto 1/N^{a}$.
The exponent $a$ is rather small and depends on the selected level (e.g., for the displayed level at 15\,\% of the spectrum we obtain $a\=0.008$, while for the level at 20\,\% we would have $a\=0.01$).
Note that the entanglement entropy for excited states for large $\lambda$ saturates at a value close to $S\approx\ln(0.482\,d_{\rm max})/\ln d_{\rm max}$, where $d_{\rm max}\!\equiv\!d(j,2j)\=2j\+1$ is the dimension of the $M\=2j$ subspace.
This expression follows from the use of the random matrix theory for estimating an average wave-function entropy for $d_{\rm max}$ orthonormal states in the limit of strong mixing \cite{Cej98}.
Note also that the critical coupling $\lambda'_c$ of the ground-state \QPT\ coincides with the sharpest increase of the ground-state entanglement entropy (Fig.\,\ref{f_entroTC}a).

\subsection{Atom--field entanglement: the $\delta\>0$ case}
\label{AFE1}

Now we are ready to turn our attention to the non-integrable $\delta\>0$ model.
Figure~\ref{f_Sdelta} shows the evolution of the entropic spectrum with parameter $\delta\in[0,1]$ for  $\omega\=\omega_0$ and $\lambda\=2.5$, well above the critical coupling $\lambda_c$ for any $\delta$.
The vertical lines demarcate the \ESQPT\ energies, the zones between them corresponding to the \D, \TC, \N\ and \S\ quantum phases. 
One can see that the departure from integrability (an increase of $\delta$) gradually destroys regular patterns in the entropic spectra, in analogy with Peres lattices.
Orderly patterns in the entropic spectrum occur in the domains of regular dynamics, while disorderly scattered points indicate chaotic domains.
Note that the accumulation of points in a narrow band close to (slightly below) the maximal entropy $S\!\approx\!1$ for high energies is a consequence of ergodicity of chaotic dynamics \cite{Str09}: a common eigenstate in the chaotic part of the spectrum contains virtually all atom--field configurations, whose dominant part is strongly entangled.
On the other hand, the regular low-energy part of the entropic spectrum tends to a minimal value $S\approx\ln 2/\ln(2j\+1)$, corresponding to states with a good parity quantum number \cite{Lam04}.

\begin{figure}[t!]
\includegraphics[width=\linewidth]{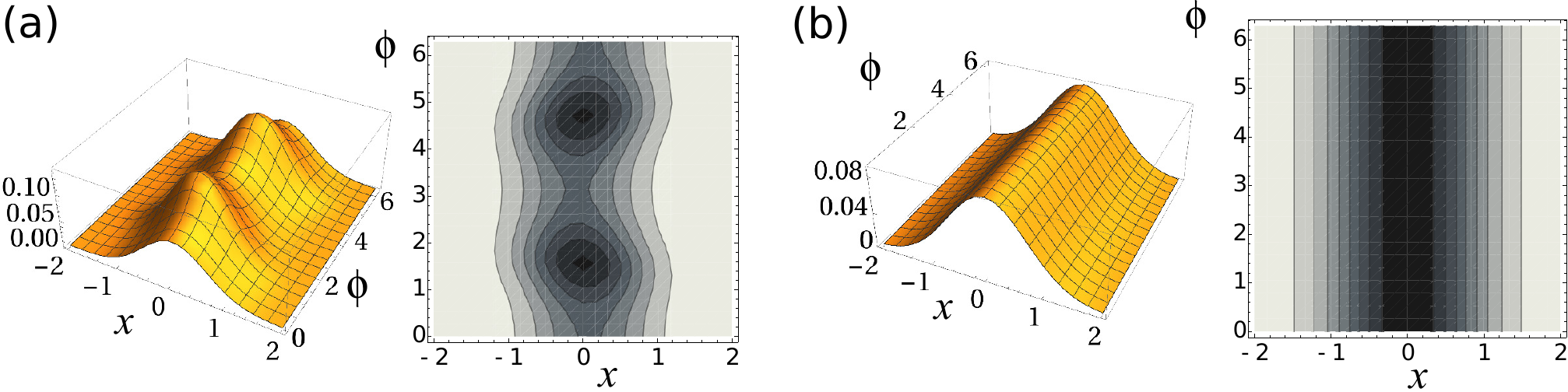}
\caption{(Color online) A detailed view of wave functions for (a) the eigenstate closest to the transition between the \TC\ and \N\ phases at $\lambda\!=\!2.5$, $\delta\!=\!0.3$ (cf.\,Fig.\,\ref{f_wav}e) and (b) the factorized $E\!=\!E_{c2}$ eigenstate for $\delta\!=\!0$. The other parameters are the same as in Fig.\,\ref{f_Sdelta}.}
\label{f_wavdet}
\end{figure} 

An interesting aspect of the entropic spectra in Fig.\,\ref{f_Sdelta} is the evolution of the atom--field entanglement in transition between the superradiant and normal phases at the critical energy $E\=-\omega_0j$.
For $\delta\>0$, this energy corresponds to $E_{c1}$ for $\lambda\in(\lambda_c,\lambda_0]$ and to $E_{c2}$ for $\lambda\in(\lambda_0,\infty)$.
We know that in the $\delta\=0$ limit, there always exists the fully factorized trivial eigenstate with $M=0$ located right at this energy and a group of weakly entangled eigenstates with low values of $M$ located nearby (see Figs.\,\ref{f_pileup} and \ref{f_detun}).
To what extent do these structures survive an increase of parameter $\delta$\,?
The answer can be read out from Fig.\,\ref{f_Sdelta}.
We see that the eigenstates with lowered entropy and even a single state with $S\!\approx\!0$ are preserved in the entropic spectra as far as the \TC\ phase is present for the chosen value of the coupling strength $\lambda$.
This is so if $\lambda\>\lambda_0$, hence $\delta\<1\-\sqrt{N\omega\omega_0/2j\lambda^2}$, see Eq.\,\eqref{lamc0scal}.
If $\lambda\in[\lambda_c,\lambda_0]$, that is for $\delta$ large enough to avoid the existence of the \TC\ phase for a given $\lambda$, the decrease of entropy no longer takes place.
These observations allow us to say that the occurrence of states with lowered atom--field entanglement entropy is a signature of the \ESQPT\ between the \N\ and \TC\ phases, but not of that between the \N\ and \D\ phases.

The wave function corresponding to the $E\!\approx\!E_{c2}$ eigenstate with the lowest entanglement entropy for $\delta\=0.3$ can be seen in Fig.\,\ref{f_wav}e.
Its detail is depicted in panel (a) of Fig.\,\ref{f_wavdet}, in comparison with the fully factorized state with $M\=0$ shown in panel (b).
Although the fine structures of the $\delta\>0$ wave function apparently prevent its full factorization (in contrast to the $\delta\=0$ case), a distant view shows a great deal of similarity to the $M\=0$ state.

Following our findings for $\delta\=0$, one might expect an analogous decrease of the atom--field entanglement entropy also at the critical energy $E_{c3}$ (in transition between the \S\ and \N\ phases) in a detuned ($\omega\!\neq\!\omega_0$) system with $\delta\>0$.
A numerical evidence of this phenomenon is however hindered by the above-presented (cf.\, the inset in Fig.\,\ref{f_entroTC}b) slowness of the decrease of the entanglement entropy dip to zero.
To see an effect in the $\delta\>0$ system, one would have to perform a truncated-space diagonalization of the Hamiltonian for a very large $j,N$ values, which is computationally demanding.

\subsection{Atom--atom entanglement}
\label{AAE}

\begin{figure}[t!]
\includegraphics[width=0.6\linewidth]{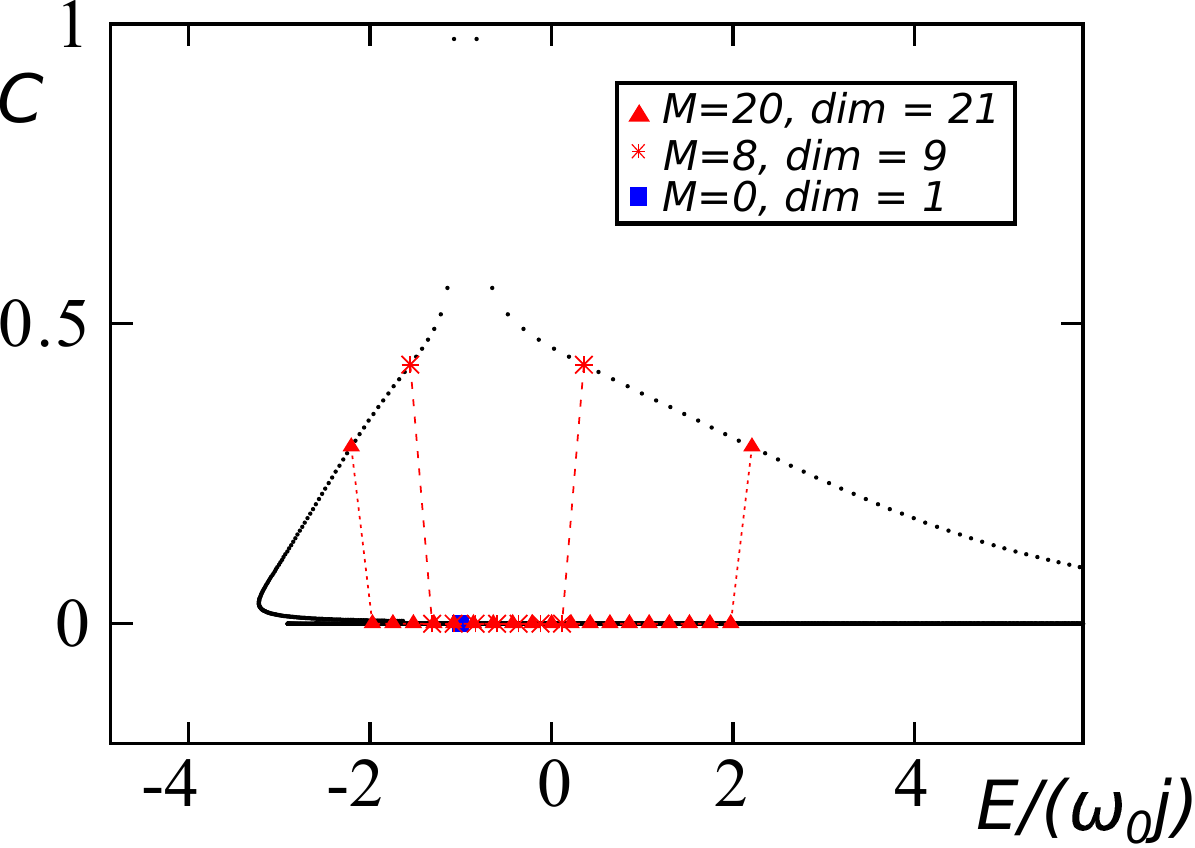}
\caption{(Color online) The scaled concurrence for the $\delta\!=\!0$ model with $\omega\!=\!\omega_0\!=\!1$, $\lambda\!=\!2.5$, $j\!=\!N/2$, $N\!=\!40$. The contribution of selected $M$-subspaces (the U-shaped chains of points) is highlighted.}
\label{f_concint}
\end{figure}
\begin{figure}[ht!]
\includegraphics[width=0.7\linewidth]{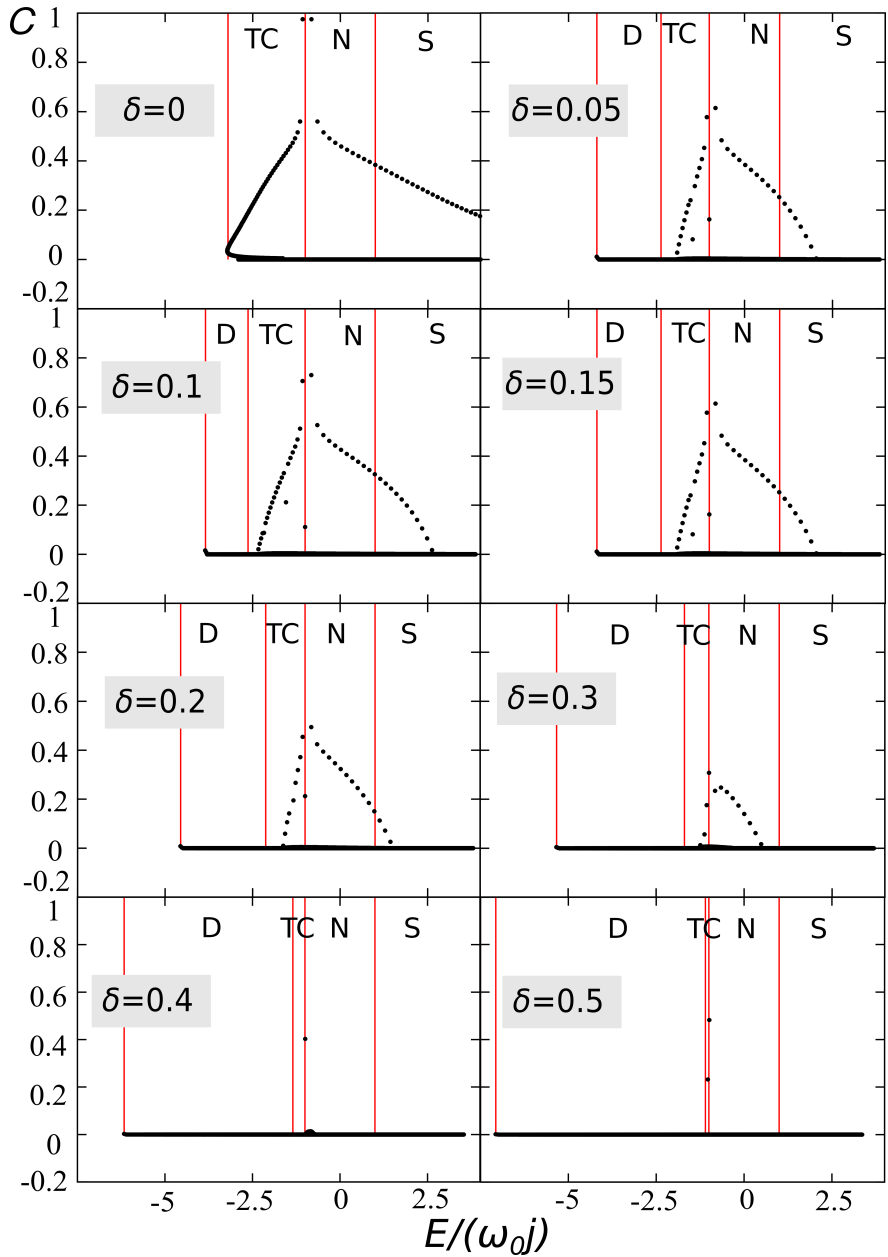}
\caption{An evolution of the atom--atom entanglement with increasing value of parameter $\delta$ for $\lambda\!=\!2.5$, $\omega\!=\!\omega_0\!=\!1$, $j\!=\!N/2$, $N\!=\!40$. The concurrence spectra contain 5000 well converged states.}
\label{f_Cdelta}
\end{figure}

We turn now to a brief numerical analysis of the atom--atom entanglement, which is limited to the symmetric $j\=N/2$ subspace of the atomic Hilbert space.
As in the previous case, we focus at first on the integrable Tavis-Cummings model.
The scaled concurrence \eqref{C} for individual eigenstates (the concurrence spectrum) of the tuned Hamiltonian with $\delta\=0$ and $\lambda\>\lambda_c$ is plotted in Fig.\,\ref{f_concint}.
The highlighted points denote states belonging to selected $M$-subspaces.
Like in the entropic spectrum (Fig.\,\ref{f_pileup}), the states corresponding the same $M$-subspace form a pattern which is always symmetric with respect to the reflection around the central energy $E(0)$.
Trivially, the $M\=0$ state $\ket{m\=-j}_{\rm A}\ket{n\=0}_{\rm F}$ has zero concurrence as the atomic state with minimal quasi-spin projection is fully factorized (all atoms in the lower level).
So this state shows no entanglement in either atom--field or atom--atom sense.
In contrast, some states with low values of $M$, which all have only weak (increasing with $M$) atom--field entanglement, show relatively large (decreasing with $M$) atom--atom entanglement.

Consider as an example a doublet of states with $M\=1$.
The scaled concurrence for these states is $C\=(N\-1)/N$, which changes between $C\=1/2$ for $N\=2$ and the maximal value $C\=1$ for $N\to\infty$ (see the upper pair of points in Fig.\,\ref{f_concint}).
As follows from Eq.\,\eqref{Hamoun}, the $M\=1$ eigenstates are expressed by the superpositions $\propto\ket{m\=-j}_{\rm A}\ket{n\=1}_{\rm F}\pm\ket{m\=-j\+1}_{\rm A}\ket{n\=0}_{\rm F}$, whose first term is fully factorized in the atomic subspace (as in the $M\=0$ case) while the second term is maximally entangled (a symmetric state with one atom in the upper level and the rest of atoms in the lower level). 
It turns out that for increasing $M\>1$, only the states with the largest positive and negative slopes within the given $M$-subspace yield a non-vanishing atom--atom entanglement.
These states form the \uvo{antennae}\ of the U-shaped dependences of $C$ in each $M$-subspace.
The scaled concurrence of these states decreases with $M$, forming together the left and right chains of $C\>0$ points in Fig.\,\ref{f_concint}.
The rest of the $\delta\=0$ eigenstates has $C\=0$. 

Figure~\ref{f_Cdelta} shows an evolution of the concurrence spectrum with increasing $\delta$ for a fixed $\lambda\>\lambda_{c}$.
Apparently, an overall trend of the atom--atom entanglement in individual eigenstates is a decrease with increasing $\delta$.
We stress that the concurrence of an absolute majority of states for any $\delta$ is zero (the spectrum contains 5000 states) and both the number of states with $C\>0$ and the values of $C$ in these states further decrease with $\delta$.
The only states in which the atom--atom entanglement remains significant even for a relatively large values of $\delta$ appear in a narrow energy interval around the \ESQPT\ critical energy $E_{c2}$ demarcating the transition between the \TC\ and \N\ phases.
These states partly coincide with those in which we previously observed a reduced atom--field entanglement, see Fig.\,\ref{f_Sdelta}.
For $\delta$ large enough to avoid the existence of the \TC\ phase, the atom--atom entanglement of all states vanishes. 

\begin{figure}[t!]
\includegraphics[width=0.9\linewidth]{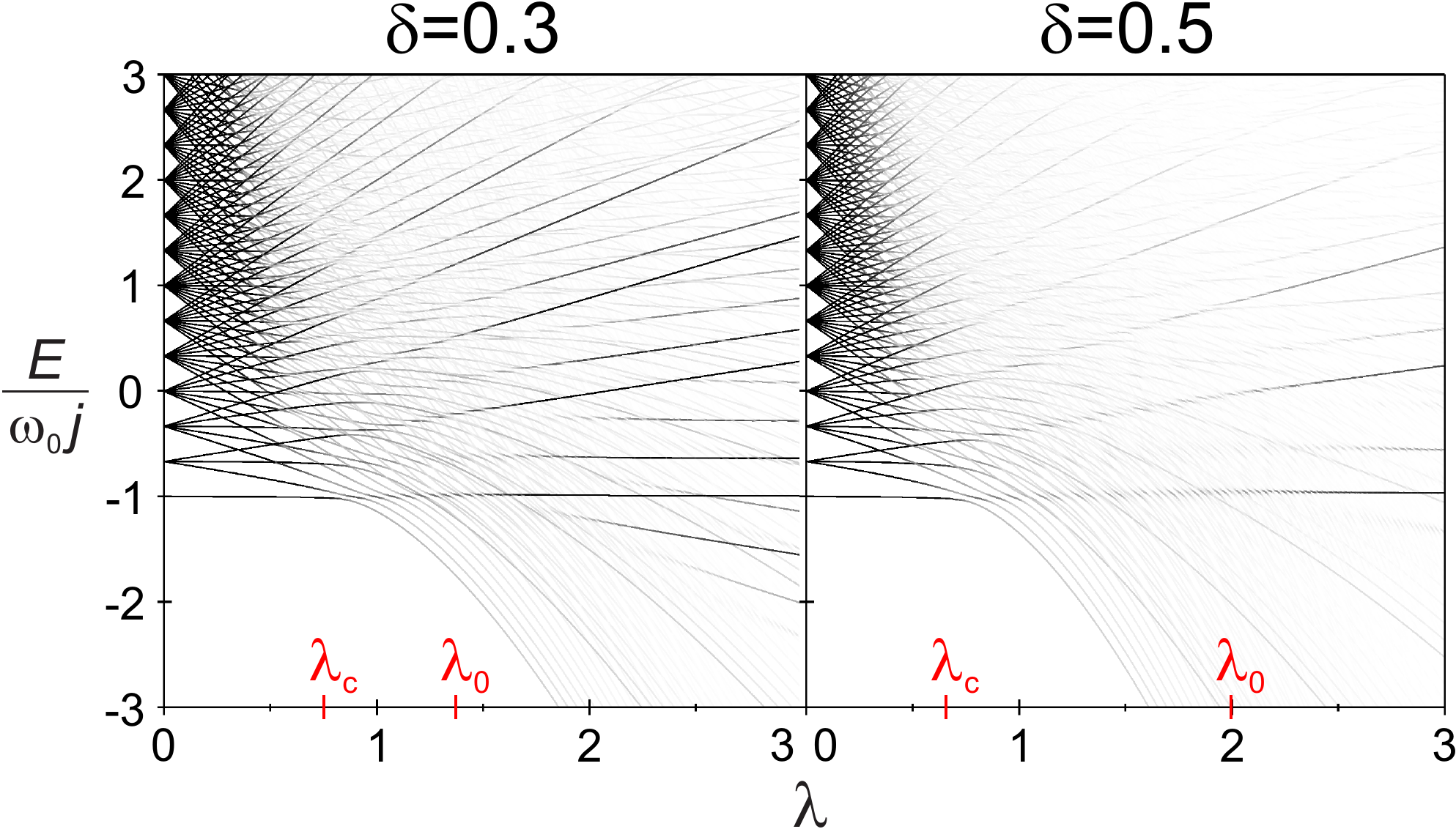}
\caption{Spectra of even-parity states of the tuned $j\!=\!N/2$ model with $N\!=\!12$ for two values of $\delta$. The number of principal components of an actual eigenstate in the unperturbed basis is encoded in the shades of gray of the respective line (black means perfect localization, $\nu_i\!=\!1$, white complete delocalization, $\nu_i\!\to\!\infty$). Revival of some $\lambda\!<\!\lambda_c$ eigenstates at $\lambda\!\sim\!\lambda_0$ is seen in both spectra.}
\label{f_fog}
\end{figure}

The states around the critical energy $E_{c2}$ with anomalous atom--field and atom--atom entanglement properties form only a small subset of all states, their appearance is nevertheless surprising.
To make this clear, let us look at these states from a slightly different perspective.
So far we have fixed a sufficiently large value of $\lambda$ and changed $\delta$, but now let us do it the other way round.
If $\delta$ is fixed at a finite value and $\lambda$ increases from 0 above $\lambda_c$ and $\lambda_0$, the states with anomalous entanglement first (not later than at $\lambda_c$) disappear from the spectrum, but above the critical coupling (somewhere in the $\lambda\sim\lambda_0$ region) some of them reappear again!
In fact, a systematic analysis of the spectrum discloses {\em approximate revivals\/} of a number of weak-coupling ($\lambda\!\lesssim\!\lambda_c$) eigenstates (not only those with anomalous entanglement) in the strong-coupling ($\lambda\!\gtrsim\!\lambda_0$) regime.
This is illustrated in Fig.\,\ref{f_fog} where we again show spectra of the tuned model with $j\=N/2$ for two values of $\delta$. 
The fullness (shade of gray) of each level encodes the number of principal components
\begin{equation}
\nu_i=\frac{1}{\sum_{i'}|\scal{\psi_{i'}(0_+)}{\psi_{i}(\lambda)}|^4}
\end{equation}
of the corresponding Hamiltonian eigenstate $\ket{\psi_{i}(\lambda)}$ in the basis of Hamiltonian eigenstates $\ket{\psi_{i'}(0_+)}$ taken at $\lambda\=0\+d\lambda\equiv 0_+$ infinitesimally displaced from the degeneracy point $\lambda\=0$ (this basis is therefore different from the $\ket{m}_{\rm A}\ket{n}_{\rm F}$ basis used above in the evaluation of the wave-function entropy).
The number of principal components varies from $\nu_i\=1$ for a perfectly localized state (identical with one of the $\lambda\=0_+$ eigenstates) to $\nu_i\!\to\!{\rm dim}{\cal H}\=\infty$ for totally delocalized states.
The revival of several $\lambda\=0_+$ eigenstates is seen in Fig.\,\ref{f_fog} as reappearance of certain dark lines in the spectrum for large $\lambda$ in the \TC\ and \N\ phases.
The mechanism underlying this phenomenon is unknown.

\section{Conclusions}
\label{SUM}

We have analyzed properties of a generalized Dicke model of single-mode superradiance allowing for a continuous (governed by parameter $\delta$) crossover between integrable Tavis-Cummings and partly chaotic Dicke limits.
We have considered three versions of the model, differing by constraints on the Hilbert space ${\cal H}\={\cal H}_{\rm A}\otimes{\cal H}_{\rm F}$.
In particular: 
(i) the all-$j$ model with the entire atomic space ${\cal H}_{\rm A}$ and $f\=N\+1$ classical degrees of freedom, 
(ii) a single-$j$ model with a single quasispin subspace ${\cal H}^{j,l}_{\rm A}\subset{\cal H}_{\rm A}$ and $f\=2$, and
(iii) a single-$j,M$ model taken for $\delta\=0$ in a subspace ${\cal H}_{M}\subset{\cal H}_{\rm A}^{j,l}\otimes{\cal H}_{\rm F}$; in this case $f\=1$.

Our first aim was to determine all types of thermal and quantum phase transitions and to characterize various phases of the atom--field system for intermediate values of $\delta$.
We have shown that the thermal phase diagram of the all-$j$ model exhibits a coexistence of the \TC\ and \D\ types of superradiant phase (with and without saddles in the free energy landscape), which contract to a single type \TC\ or \D\ in the limits $\delta\=0$ or 1, respectively.
The quantum phase diagram of a single-$j$ model contains the superradiant \QPT\ and three types of \ESQPTs, characterized by singularities (jumps and a logarithmic divergence) in the first derivative of the semiclassical level density as a function of energy.
The quantum phase diagram of a critical single-$j,M$ model with $M\=2j$ shows for $\omega\>\omega_0$ another second-order \QPT\ and for $\omega\!\neq\!\omega_0$ also another \ESQPT.
The latter is characterized by a singularity (logarithmic divergence) directly in the level density in the $M\=2j$ subspace.

We have associated four spectral domains in between the \ESQPT\ critical borderlines of the single-$j$ model with the \TC, \D, \N\ and \S\ quantum phases of the model.
These phases are characterized by different shapes of a smoothed energy dependence of some expectation values, which show abrupt changes at the \ESQPT\ energies.
A natural choice of the phase-defining quantity is the interaction Hamiltonian, whose expectation value $\ave{H_{\rm int}}_i$ determines the slope $dE_i/d\lambda$ of the given level.
According to the density--flow relation of the \ESQPT\ signatures \cite{Str16}, a smoothed level slope should exhibit the same type of non-analyticity as the semiclassical level density.

Our second aim was to analyze atom--field and atom--atom entanglement properties of the model in a wide excited domain.
This was first done in the integrable single-$j,M$ model, which allowed for a qualitative explanation of results, and then numerically in the single-$j$ model with $j\=N/2$.
We have found that an absolute majority of eigenstates in the spectrum for any choice of model parameters has large atom--field entanglement but vanishing atom--atom entanglement.
Exceptional in this sense for $\delta\=0$ are the eigenstates with low values of $M$, which yield a weak atom--field entanglement and simultaneously an increased atom--atom entanglement.
We showed that remnants of these states exist relatively far in the non-integrable regime with $\delta\>0$.
This concerns in particular the state with $M\=0$ located at the \ESQPT\ energy $E_{c2}$, which has zero atom--field entanglement. 
Another state with reduced atom--field entanglement entropy, originating in the critical $M\=2j$ subspace, appears at the \ESQPT\ energy $E_{c3}$ for $\omega\!\neq\!\omega_0$.

Although the above-discussed states with anomalous entanglement properties have been located near the \ESQPT\ critical borderlines, we do not regard this connection as systematic.
The \ESQPT\ singularity in the spectrum has strong consequences on the structure of eigenvectors---hence possibly also on their entanglement properties---in the $f\=1$ case, as indeed observed in the $M\=2j$ subspace of the present model.
However, for $f\>1$ it does not seem likely that the entanglement plays a substantial role in a generic \ESQPT\ since transitions between quantum phases affect (as we have seen) the trends of eigenstate variations rather than the eigenstates themselves.
Nevertheless, it should be stressed that robust, systematic studies of entanglement in excited states are still rather scarce.
We hope that results of our analysis will initiate similar studies in other relevant systems.

\section*{Acknowledgments}
We acknowledge discussions with T.\,Brandes, N.\,Lambert, J.\,Hirsch and M.A.\,Bastarrachea-Magnani, and support of the Czech Science Foundation, project no.\,P203-13-07117S.


\end{document}